\documentclass[10pt]{article}

\usepackage{amsmath,amssymb,epsfig,color,bbold}
\usepackage{feynmp-auto}
\usepackage[hidelinks]{hyperref}
\usepackage{caption}
\usepackage{subcaption}
\usepackage{slashbox}
\usepackage{physics}
\newcommand{\marrow}[5]{%
    \fmfcmd{style_def marrow#1
    expr p = drawarrow subpath (1/4, 3/4) of p shifted 6 #2 withpen pencircle scaled 0.4;
    label.#3(btex #4 etex, point 0.5 of p shifted 6 #2);
    enddef;}
    \fmf{marrow#1,tension=0}{#5}}

\newcommand{\Noarrow}[6]{%
    \fmfcmd{style_def marrow#1
    expr p = drawarrow subpath (1/2, 1/2) of p shifted #6 #2 withpen pencircle scaled 0.4;
    label.#3(btex #4 etex, point 0.5 of p shifted #6 #2);
    enddef;}
    \fmf{marrow#1,tension=0}{#5}}

\usepackage{psfrag}
\usepackage{graphicx,xspace}
\usepackage{bm}

\usepackage[affil-it, auth-sc]{authblk}
\usepackage{lineno}
\usepackage[sort&compress,numbers]{natbib}
\usepackage{doi}
\usepackage{eso-pic}
\usepackage[capitalise]{cleveref}

\newcommand{\be}{\begin{equation}}  
\newcommand{\ee}{\end{equation}} 
\def\slash#1{#1\!\!\!/\!\,\,}  
\newcommand{\nl}{\nonumber \\ }

\renewcommand{\order}{{\cal O}}
\long\def\symbolfootnote[#1]#2{\begingroup%
\def\thefootnote{\fnsymbol{footnote}}\footnote[#1]{#2}\endgroup}

\def\qMark{\textsf{?}\xspace}
\def\dd{\mathrm{d}} 
\def\e{\mathrm{e}}

\allowdisplaybreaks

\begin{document}

\begin{fmffile}{feynmffile} 
\fmfcmd{%
vardef middir(expr p,ang) = dir(angle direction length(p)/2 of p + ang) enddef;
style_def arrow_left expr p = shrink(.7); cfill(arrow p shifted(4thick*middir(p,90))); endshrink enddef;
style_def arrow_left_more expr p = shrink(.7); cfill(arrow p shifted(6thick*middir(p,90))); endshrink enddef;
style_def arrow_right expr p = shrink(.7); cfill(arrow p shifted(4thick*middir(p,-90))); endshrink enddef;}

\fmfset{arrow_ang}{15}
\fmfset{arrow_len}{2.5mm}
\fmfset{decor_size}{3mm}
\AddToShipoutPictureFG*{\AtPageUpperLeft{\put(-60,-75){\makebox[\paperwidth][r]{FERMILAB-PUB-24-0070-T}}}}
\AddToShipoutPictureFG*{\AtPageUpperLeft{\put(-60,-60){\makebox[\paperwidth][r]{CALT-TH-2024-006}}}}


\title{\Large\bf Renormalization of beta decay at three loops and beyond}
\author[1,2]{Kaushik~Borah}
\author[1,2]{Richard~J.~Hill}
\author[3]{Ryan~Plestid}
\affil[1]{University of Kentucky, Department of Physics and Astronomy, Lexington, KY 40506 USA \vspace{1.2mm}}
\affil[2]{Fermilab, Theoretical Physics Department, Batavia, IL 60510 USA
\vspace{1.2mm}}
\affil[3]{Walter Burke Institute for Theoretical Physics, \\
California Institute of Technology, Pasadena, CA 91125 USA\vspace{1.2mm}}

\date{\today}

\maketitle

\begin{abstract}
  \vspace{0.2cm}
  \noindent
The anomalous dimension for heavy-heavy-light effective theory operators describing nuclear beta decay is computed through three-loop order in the static limit. The result at order $Z^2\alpha^3$ corrects a previous result in the literature.  An all-orders symmetry is shown to relate the anomalous dimensions at leading and subleading powers of $Z$ at a given order of $\alpha$. The first unknown coefficient for the anomalous dimension now appears at $O(Z^2\alpha^4)$. 
\end{abstract}
\vfil

\newpage
\tableofcontents
\newpage

\section{Introduction}

Precision measurements of nuclear beta decay rates provide important constraints on fundamental constants~\cite{Bopp:1986rt,Ando:2004rk,Darius:2017arh,Seng:2018yzq,Seng:2018qru,Fry:2018kvq,Czarnecki:2019mwq,Hayen:2020cxh,Seng:2020wjq,Gorchtein:2021fce,Hardy:2020qwl,UCNt:2021pcg,Shiells:2020fqp,Cirigliano:2022hob} 
and can probe new physics~\cite{Cirigliano:2013xha,Glick-Magid:2016rsv,Gonzalez-Alonso:2018omy,Glick-Magid:2021uwb,Falkowski:2021vdg,Brodeur:2023eul,Crivellin:2020ebi,Coutinho:2019aiy,Crivellin:2021njn,Crivellin:2020lzu,Cirigliano:2022yyo}. 
Quantum electrodynamics (QED) radiative corrections 
are enhanced by the charge $Z$ of the nucleus, and must be systematically incorporated. For example, extractions of the Cabibbo–Kobayashi–Maskawa (CKM) matrix element $|V_{ud}|$ from the inclusive decay rates of super-allowed beta decays require a careful account of both short- and long-distance radiative corrections \cite{Seng:2018yzq,Seng:2018qru,Czarnecki:2019mwq,Hardy:2020qwl,Hill:2023acw}. Similarly, searches for the energy-dependent  Fierz Interference term \cite{Cirigliano:2013xha,Glick-Magid:2016rsv,Brodeur:2023eul}, which would shed light on currents induced by short-distance physics beyond the Standard Model \cite{Cirigliano:2013xha,Falkowski:2020pma,Cirigliano:2023nol},  requires a detailed understanding of the energy dependence of the beta decay spectrum and therefore a detailed understanding of energy dependent QED radiative corrections \cite{Hayen:2017pwg}. 

Radiative corrections receive contributions from many scales, ranging from the weak scale $\mu \sim M_W$, to the hadronic scale $\mu\sim m_p$ and scales sensitive to nuclear structure $\mu \sim R^{-1}$. 
Physics at short distance scales can be integrated out, 
to obtain an effective field theory (EFT) that describes scales of order 
$\mu \sim Q \sim m_e$ with $Q=\Delta M$ the mass splitting between the parent and daughter nucleus and $m_e$ the electron mass \cite{Hill:2023acw}. For a nuclear beta decay $A \rightarrow B \nu e$,  
the effective theory Lagrangian is 
\begin{align}  
    \label{eff-L}
    {\cal L} &= {\cal L}_{\rm QED} + {\cal L}_\nu +  \bar{h}_v^{(A)} \left( iv\cdot \partial + e Q_A v\cdot A \right) h_v^{(A)} 
    + 
\bar{h}_v^{(B)} \left( iv\cdot \partial + e Q_B v\cdot A \right) h_v^{(B)} 
+ {\cal C}\, \bar{h}_v^{(B)} \Gamma h_v^{(A)} \, \bar{\nu} \Gamma^\prime e \,,
\end{align}
where ${\cal L}_{\rm QED}$  is the QED Lagrangian for a single electron field, and ${\cal L}_\nu$ contains the kinetic term for the neutrino. 
Beta decay is mediated by a contact operator with a Wilson coefficient, ${\cal C} \sim G_F$,  and $v^\mu = (1,0,0,0)$ is the four-velocity of the nucleus $A$ and nucleus $B$ (we work in the static limit neglecting nuclear recoil such that $v^\mu_A=v^\mu_B=v^\mu$). All short-distance physics, including nuclear structure, has been integrated out and absorbed into the Wilson coefficient $\mathcal{C}$. We focus on the theory at leading power, i.e. we consider terms that are finite in the limit 
$pR\rightarrow 0$ with $p$ the momentum of the electron in the lab frame.

The effective theory (\ref{eff-L}) 
separates contributions into a low-energy matrix element $\mathcal{M}(\mu_L)$ that can be evaluated at the ``low-scale'' $\mu_L\sim Q \sim m_e$, 
and the Wilson coefficient of operators at the same scale ${\cal C}(\mu_L)$. The Wilson coefficient at the low scale contains logarithmic enhancements, $\log(\mu_H/\mu_L)$, due to renormalization group (RG) evolution from the ``high-scale'', $\mu_H\sim R^{-1}$ (where matching from a theory with explicit nuclear structure is performed), down to the low-scale, $\mu_L$. The RG evolution is described by
\begin{equation}\label{eq:RG}
    {\cal C}(\mu_L) = {\cal C}(\mu_H) \exp\left[- \int_{\alpha_L}^{\alpha_H} \dd \alpha ~\frac{\gamma(\alpha)}{\beta(\alpha)} \right] ~,
\end{equation}
where $\gamma$ is the anomalous dimension of the operator associated to the Wilson coefficient of interest, $\beta=d\alpha/d\log\mu$ 
determines the running of the QED coupling, and $\alpha_{L,H} = \alpha(\mu_{L,H})$. To systematically resum these logarithmic enhancements, we need the anomalous dimension for the effective operator multiplying $\mathcal{C}$ in Eq.~(\ref{eff-L}).  

In what follows we present the formalism for the heavy-heavy-light EFT operators in \cref{eff-L} describing long-distance QED radiative corrections to nuclear beta decay~\cite{Hill:2023acw}.  
We compute the anomalous dimension of these operators through order $Z^2\alpha^3$. 
In the massless-electron limit we show that a symmetry relates different powers of $Z$ at the same order in $\alpha$.  In particular, this symmetry relates the anomalous dimension at order $Z\alpha^3$ to that at order $Z^2\alpha^3$, and the anomalous dimension at order $Z^3\alpha^4$ to that at order $Z^4\alpha^4$. Since we compute the $Z^2\alpha^3$ result, and the $Z^4\alpha^4$ result is known via a solution of the Dirac equation with a static background Coulomb field \cite{Hill:2023bfh},  this implies that the first unknown coefficient now appears at order $Z^2\alpha^4$.

The remainder of the paper is structured as follows.  In 
Sec.~\ref{sec:symm} we describe the new symmetry emerging in the massless electron limit. 
Section~\ref{sec:constants} introduces notation for amplitudes and renormalization constants.  
Section~\ref{sec:eval} evaluates amplitudes at one, two and three-loop order, and Sec.~\ref{sec:anom} extracts anomalous dimension coefficients. 
Section~\ref{sec:disc} provides a summary discussion. 
Appendices provide details on the background field Feynman rules used in the main text, 
and on the evaluation of three-loop integrals. 

\vfill
\pagebreak

\section{Effective operators and symmetry \label{sec:symm}}

Consider 
the heavy-heavy-light effective operator
\begin{align}\label{eq:Oeff}
O_{IJk} &=   \big( \bar{h}_v^{[Z]} \big)_I  
\big( h_v^{[Z+1]} \big)_J \,\, e_k 
\end{align}
in QED with one massless electron.   
Here $h_v^{[Z]}$ is a heavy particle field of electric charge $Z$, and  $v^\mu=(1,0,0,0)$ in the rest frame of the heavy particle;   
 $e$ is the relativistic electron field of electric charge $-1$.  
The anomalous dimensions of interest are diagonal in the spin indices $I,J$ for the heavy particles and $k$ for the electron. 
The matrix element of $O_{IJk}$ can be expanded in powers 
of the QED coupling $e^2$.  The contribution at a given order, 
$(e^2)^n$, can be further decomposed into separate, gauge invariant contributions involving different powers of $Z$:  
$Z^0, \dots, Z^{2n}$.  It is straightforward to show that $n$-loop contributions with powers of $Z$ larger than $n$ vanish \cite{Beg:1969zu,Hill:2023acw}.  

To compute the anomalous dimension of the operator (\ref{eq:Oeff}) we consider 
amplitudes involving one insertion of the operator, 
neglecting the electron mass and external momenta of all particles. 
Infrared divergences are regulated by including a photon mass $\lambda$. 
In this limit, symmetry enforces constraints relating different powers of $Z$
at the same order in $e^2$. 
In order to demonstrate the symmetry constraints,
we generalize and consider the heavy-heavy-light amplitudes 
involving massless electrons of charge $Q_e$,
initial heavy particle of charge $Q_A$ and final heavy particle of charge $Q_B$.  
The amplitudes under consideration are 
invariant under the simultaneous transformation 
$Q_e \to -Q_e$ and $Q_A \leftrightarrow Q_B$. 
We are interested in the particular case $Q_e=-1$, $Q_A=Z+1$ and $Q_B=Z$.   

Consider the amplitude with arbitrary photon 
attachments to the electron 
(with polarizations $\mu_1,\mu_2,\dots$ and incoming momenta $L_1, L_2, \dots$),
to the heavy particle $A$ (with polarization $\nu_1, \nu_2, \dots$
and momenta $K_1,K_2,\dots$) and to the heavy particle $B$
(with polarization $\rho_1,\rho_2,\dots$ and momenta $P_1,P_2,\dots$). 
Using heavy-particle Feynman rules, the amplitude is
\begin{align}
\nl
&    \parbox{70mm}{
\begin{fmfgraph*}(200,150)
\fmfstraight
  \fmfleftn{l}{5}
  \fmfrightn{r}{5}
  \fmfbottomn{b}{9}
  \fmftopn{t}{9}
  \fmf{phantom}{l2,v,r2}
  \fmf{phantom}{l4,v,r4}
  \fmffreeze
  \fmf{phantom}{r4,e1,e2,e3,v}
  \fmf{phantom}{l2,A1,A2,A3,v}
  \fmf{phantom}{r2,B1,B2,B3,v}
  \fmffreeze
  \fmf{fermion}{r4,e1,e2,v}
  \fmf{double}{l2,v}
  \fmf{double}{r2,v}
   \fmfv{decor.shape=circle,decor.filled=shaded}{v}
   \fmffreeze
   \fmf{photon}{t8,e1}
   \fmf{photon}{t7,e2}
   \fmf{photon}{b2,A1}
   \fmf{photon}{b3,A2}
   \fmf{photon}{b8,B1}
   \fmf{photon}{b7,B2}
    \marrow{a}{right}{rt}{$L_1$}{t8,e1}
    \marrow{b}{right}{rt}{$L_2$}{t7,e2}
    \marrow{c}{left}{lft}{$K_1$}{b2,A1}
    \marrow{d}{left}{lft}{$K_2$}{b3,A2}
    \marrow{e}{right}{rt}{$P_1$}{b8,B1}
    \marrow{f}{right}{rt}{$P_2$}{b7,B2}
    \fmflabel{$\mu_1$}{t8}
    \fmflabel{$\mu_2$}{t7}
    \fmflabel{$\nu_1$}{b2}
    \fmflabel{$\nu_2$}{b3}
    \fmflabel{$\rho_1$}{b8}
    \fmflabel{$\rho_2$}{b7}
    \Noarrow{g}{left}{lft}{$...$}{b4,A3}{0}
    \Noarrow{h}{right}{rt}{$...$}{b6,B3}{0}
    \Noarrow{i}{right}{rt}{$...$}{t6,e3}{0}
\end{fmfgraph*}
}
\nl
\nl
&= \Bigg[ \cdots {(-e Q_e)(\slash{L}_1+\slash{L}_2)\gamma^{\mu_2} \over (L_1+L_2)^2+i0}
{(-e Q_e) \slash{L}_1 \gamma^{\mu_1} \over L_1^2+i0}
\Bigg]
\Bigg[ 
\cdots {-e Q_A v^{\nu_2}\over v\cdot(K_1+K_2)+i0}
{-e Q_A v^{\nu_1} \over v\cdot K_1+i0}
\Bigg]\times
\nl
&\quad \times
\Bigg[ 
\cdots {-e Q_B v^{\rho_1}\over v\cdot (-P_1)+i0}
{-e Q_B v^{\rho_2} \over v\cdot (-P_1 - P_2) + i0}
\Bigg] \,, 
\nl
\end{align}
where we have suppressed the indices $IJk$ from Eq.~\ref{eq:Oeff} 
and the external spinor wavefunctions.  
It is readily seen that under the transformation $Q_e \to -Q_e$, 
$Q_A \leftrightarrow Q_B$, $L_i \to -L_i$, $K_i \to -K_i$ and $P_i \to -P_i$, we have 
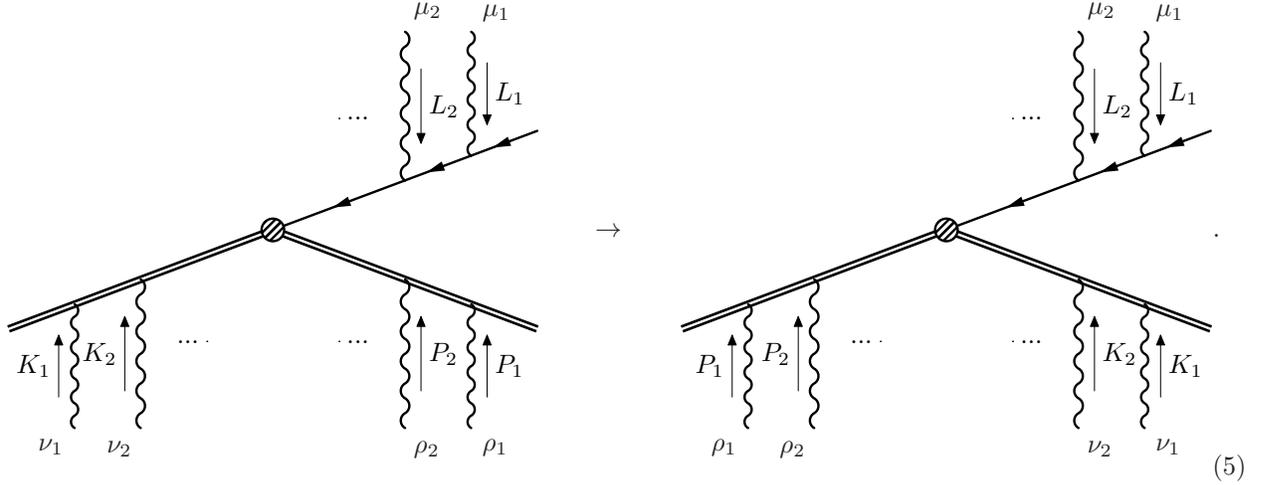
\begin{align}
\nl
    \parbox{70mm}{
\begin{fmfgraph*}(200,150)
\fmfstraight
  \fmfleftn{l}{5}
  \fmfrightn{r}{5}
  \fmfbottomn{b}{9}
  \fmftopn{t}{9}
  \fmf{phantom}{l2,v,r2}
  \fmf{phantom}{l4,v,r4}
  \fmffreeze
  \fmf{phantom}{r4,e1,e2,e3,v}
  \fmf{phantom}{l2,A1,A2,A3,v}
  \fmf{phantom}{r2,B1,B2,B3,v}
  \fmffreeze
  \fmf{fermion}{r4,e1,e2,v}
  \fmf{double}{l2,v}
  \fmf{double}{r2,v}
   \fmfv{decor.shape=circle,decor.filled=shaded}{v}
   \fmffreeze
   \fmf{photon}{t8,e1}
   \fmf{photon}{t7,e2}
   \fmf{photon}{b2,A1}
   \fmf{photon}{b3,A2}
   \fmf{photon}{b8,B1}
   \fmf{photon}{b7,B2}
    \marrow{a}{right}{rt}{$L_1$}{t8,e1}
    \marrow{b}{right}{rt}{$L_2$}{t7,e2}
    \marrow{c}{left}{lft}{$K_1$}{b2,A1}
    \marrow{d}{left}{lft}{$K_2$}{b3,A2}
    \marrow{e}{right}{rt}{$P_1$}{b8,B1}
    \marrow{f}{right}{rt}{$P_2$}{b7,B2}
    \fmflabel{$\mu_1$}{t8}
    \fmflabel{$\mu_2$}{t7}
    \fmflabel{$\nu_1$}{b2}
    \fmflabel{$\nu_2$}{b3}
    \fmflabel{$\rho_1$}{b8}
    \fmflabel{$\rho_2$}{b7}
    \Noarrow{g}{left}{lft}{$...$}{b4,A3}{0}
    \Noarrow{h}{right}{rt}{$...$}{b6,B3}{0}
    \Noarrow{i}{right}{rt}{$...$}{t6,e3}{0}
\end{fmfgraph*}
}
& \quad  \quad \to \quad \quad
   \parbox{70mm}{
\begin{fmfgraph*}(200,150)
\fmfstraight
  \fmfleftn{l}{5}
  \fmfrightn{r}{5}
  \fmfbottomn{b}{9}
  \fmftopn{t}{9}
  \fmf{phantom}{l2,v,r2}
  \fmf{phantom}{l4,v,r4}
  \fmffreeze
  \fmf{phantom}{r4,e1,e2,e3,v}
  \fmf{phantom}{l2,A1,A2,A3,v}
  \fmf{phantom}{r2,B1,B2,B3,v}
  \fmffreeze
  \fmf{fermion}{r4,e1,e2,v}
  \fmf{double}{l2,v}
  \fmf{double}{r2,v}
   \fmfv{decor.shape=circle,decor.filled=shaded}{v}
   \fmffreeze
   \fmf{photon}{t8,e1}
   \fmf{photon}{t7,e2}
   \fmf{photon}{b2,A1}
   \fmf{photon}{b3,A2}
   \fmf{photon}{b8,B1}
   \fmf{photon}{b7,B2}
    \marrow{a}{right}{rt}{$L_1$}{t8,e1}
    \marrow{b}{right}{rt}{$L_2$}{t7,e2}
    \marrow{c}{left}{lft}{$P_1$}{b2,A1}
    \marrow{d}{left}{lft}{$P_2$}{b3,A2}
    \marrow{e}{right}{rt}{$K_1$}{b8,B1}
    \marrow{f}{right}{rt}{$K_2$}{b7,B2}
    \fmflabel{$\mu_1$}{t8}
    \fmflabel{$\mu_2$}{t7}
    \fmflabel{$\rho_1$}{b2}
    \fmflabel{$\rho_2$}{b3}
    \fmflabel{$\nu_1$}{b8}
    \fmflabel{$\nu_2$}{b7}
    \Noarrow{g}{left}{lft}{$...$}{b4,A3}{0}
    \Noarrow{h}{right}{rt}{$...$}{b6,B3}{0}
    \Noarrow{i}{right}{rt}{$...$}{t6,e3}{0}
\end{fmfgraph*}
} \,.
\nl
\end{align}
Since the photon propagators and loop integration measure are invariant under this transformation, the result is unchanged after summing over subamplitudes.

In conjunction with the constraint that powers of $Z$ larger than the number of loops do not appear, invariance under $Q_e\to -Q_e$ and $Q_A \leftrightarrow Q_B$ 
enforces constraints on the $Z$ dependence of the anomalous dimension at each order in perturbation theory.  
For example, at one-loop order the possible charge structures are 
\begin{align}
Q_e^2 \,, Q_e (Q_A - Q_B) 
\to 1\,, 1 \,,
\end{align}
i.e., independent of $Z$.  
At two-loop order, the possible structure are 
\begin{align}
Q_e^4 \,, Q_e^3 (Q_A-Q_B) \,, 
Q_e^2(Q_A - Q_B)^2 \,, Q_e^2 Q_A Q_B  
\to 1 \,, 1 \,, 1 \,, Z(Z+1) \,,
\end{align}
i.e., a linear combination of $1$ and $Z(Z+1)$. 
At arbitrary loop order, the amplitude can similarly be expressed as a linear combination of powers of $Q_e^2 \to 1$, $Q_e(Q_A-Q_B) \to 1$ and $Q_A Q_B \to Z(Z+1)$, with the total power of $Q_A$ and $Q_B$ not exceeding the loop order. 
It is straightforward to see that the amplitude, and thus 
the anomalous dimension at a given loop order
$n$ is a linear combination of powers $Z^i(Z+1)^i$, where $2i \le n$. 

Expanding 
\begin{align}
\gamma_{\cal O} &
= \sum_{n=0}^\infty  \left(\alpha \over 4\pi\right)^{n+1} \gamma_n 
=
\sum_{n=0}^\infty \sum_{i=0}^{n+1} \left(\alpha \over 4\pi\right)^{n+1} \gamma_n^{(i)}Z^{n+1-i} 
= \sum_{i=0}^{n+1} \gamma^{(i)}Z^{n+1-i} 
\,,
\end{align}
we have in particular at one-loop order, 
$\gamma_0 = Z \gamma_0^{(0)} + \gamma_0^{(1)}$ with 
\begin{align}
    \gamma_0^{(0)} = 0 \,.  
\end{align}
At two-loop order, 
$\gamma_1 = Z^2 \gamma_1^{(0)} + Z \gamma_1^{(1)} + \gamma_1^{(2)}$ 
with 
\begin{align}
\gamma_1^{(1)}= \gamma_1^{(0)} \,,
\end{align}
and at three-loop order, $\gamma_2 = Z^3 \gamma_2^{(0)} + Z^2 \gamma_2^{(1)} + Z \gamma_2^{(2)} + \gamma_2^{(3)}$ with 
\begin{align}
\gamma_2^{(0)} = 0 \,,\quad \gamma_2^{(2)}= \gamma_1^{(1)} \,.
\end{align}
Using the known Dirac solution~\cite{Hill:2023acw}, $\gamma^{(0)}$,
at leading power in $Z$, the number of undetermined coefficients describing $Z$ dependence of the anomalous dimension (i.e., beyond the result $\gamma_n^{(n+1)}$ for the local heavy-light current)
is zero for two loops, and one for three loops.  
Extended to four loops, the constraints imply
$\gamma_3^{(1)}=2\gamma_3^{(0)}$,
$\gamma_3^{(3)}=\gamma_3^{(2)}-\gamma_3^{(0)}$, 
leaving a single undetermined coefficient $\gamma_3^{(2)}$
(beyond the Dirac solution $\gamma_3^{(0)}$ and the heavy-light, $Z=0$, limit $\gamma_3^{(4)}$). 
At arbitrary even loop order, 
\begin{align}\label{eq:evenloopsymm}
\gamma_{2n-1}^{(1)}=n\gamma_{2n-1}^{(0)} \,,
\end{align}
and there are $n-1$ undetermined coefficients ($2n$ loops, $n=1,2,\dots$).
At arbitrary odd loop order, 
\begin{align}
\gamma_{2n}^{(2)}=n\gamma_{2n}^{(1)} \,,
\end{align}
and there are $n$ undetermined coefficients ($2n+1$ loops, $n=1,2,\dots$). 

\section{Renormalization constants \label{sec:constants}} 

\begin{figure}
\begin{center}

\vspace{5mm}
\parbox{10mm}{
\begin{fmfgraph*}(30,30)
 \fmftop{t}
 \fmfbottom{b}
 \fmf{photon}{t,b}
  \fmfv{decor.shape=cross}{b}
  \fmflabel{$\mu$}{t}
\end{fmfgraph*}
}
= $iZe \, \delta_0^\mu \, 2\pi \delta(q^0)$
\qquad
\parbox{10mm}{
\begin{fmfgraph*}(30,30)
    \fmfstraight
    \fmftop{t}
    \fmfbottomn{b}{3}
    \fmf{photon}{t,b2}
    \fmf{double}{b1,b2,b3}
    \fmflabel{$\mu$}{t}
\end{fmfgraph*}
}
= $i e \, \delta^\mu_0$ 
\qquad
\parbox{10mm}{
\begin{fmfgraph*}(30,30)
    \fmfstraight
    \fmftop{t}
    \fmfbottomn{b}{3}
    \fmf{photon}{t,b2}
    \fmf{fermion}{b1,b2,b3}
    \fmflabel{$\mu$}{t}
\end{fmfgraph*}
}
= $-i e \, \gamma^\mu$ 

\vspace{5mm}

\end{center}
\caption{\label{fig:vertex} Vertex Feynman rules.  The cross denotes insertion of background field.}
\end{figure}

\begin{figure}
\begin{center}
\vspace{5mm}
\parbox{28mm}{
\begin{fmfgraph*}(60,40)
  \fmfleftn{l}{3}
  \fmfrightn{r}{3}
  \fmf{photon}{l2,x,y,r2}
  \marrow{a}{up}{top}{$q$}{x,y}
  \fmflabel{$\mu$}{l2}
  \fmflabel{$\nu$}{r2}
\end{fmfgraph*}
}
= ${\displaystyle {-i \over q^2 -\lambda^2 + i0} \left( g_{\mu\nu} - (1-\xi) {q_\mu q_\nu \over q^2}  \right)}$
\qquad
\parbox{28mm}{
\begin{fmfgraph*}(60,40)
  \fmfleftn{l}{3}
  \fmfrightn{r}{3}
  \fmf{double}{l2,x,y,r2}
  \marrow{a}{up}{top}{$q$}{x,y}
\end{fmfgraph*}
}
= ${\displaystyle {i \over q^0 + i0}}$
\end{center}
\caption{\label{fig:prop} Propagator Feynman rules, with gauge parameter $\xi$ and photon mass $\lambda$.}
\end{figure}

To determine the anomalous dimension of the effective operator (\ref{eq:Oeff}) we compute the amputated amplitude involving one insertion of the effective operator (\ref{eq:Oeff}).  
We work in arbitrary covariant gauge with 
gauge parameter $\xi$,  
and photon mass $\lambda$ to 
regulate infrared divergences.
We use dimensional regularization to
regulate ultraviolet divergences using spacetime dimension 
$d\equiv 4-2\epsilon$, or equivalently space dimension 
$D=d-1=3-2\epsilon$. 

Let us compute the onshell amplitude for the process 
$[Z+1] \to [Z] e^+$, i.e. for initial state heavy nucleus of charge $Z+1$ and final state heavy nucleus of charge $Z$ and positron. 
It is convenient to compute the equivalent process 
$[+1] \to e^+$ in the presence of a background field 
for charge $Z$, where $[+1]$ is a heavy ``proton" field with charge $+1$. 
The Feynman rules are given in Figs.~\ref{fig:vertex} and \ref{fig:prop}.
These Feynman rules are equivalent, order by order in perturbation theory to a brute force calculation in terms of the original heavy-particle Feynman rules [i.e., those derived from Eq.~(\ref{eff-L})] for the $[Z+1] \to [Z] e^+$ process.  
However, the number of diagrams is drastically reduced, and powers of $Z$ can be easily isolated \cite{eikonal_algebra}.
As discussed in \cite{eikonal_algebra} and explicitly shown in Appendix~\ref{sec:backgroundfield}, diagrams in which the heavy ``proton" field interacts with the background field do not contribute to the anomalous dimension and can be neglected for our purposes.

In terms of the bare coupling $e_0$ (whose mass dimension is $[e_0]=2-d/2$), 
let us introduce the dimensionless quantity 
\begin{align} 
\hat{e}_0^2 = {e_0^2 \over (4\pi)^2}[(4\pi)^\epsilon \Gamma(1+\epsilon)] \lambda^{-2\epsilon} \,. 
\end{align}
The amputated amplitude can be expanded as 
\begin{align}
{\cal M} &= 1 + \hat{e}_0^2  {\cal M}_1 
 + \hat{e}_0^4 {\cal M}_2 
 + \hat{e}_0^6 {\cal M}_3 + \order(e_0^8) \,,
\end{align}
where we use that the photon mass $\lambda$ is the only mass scale in the problem.  
The one-, two- and three-loop contributions can be written as 
, 
\begin{align}\label{eq:amp}
{\cal M}_1 &= Z \left( A_{1,0} + A_{1,1}\epsilon + \dots \right)
+ \left( {B_{1,-1}\over \epsilon} + B_{1,0} + B_{1,1}\epsilon + \order(\epsilon^2) \right)
\nl
{\cal M}_2 &= Z^2 \left( {A_{2,-1}\over \epsilon} + A_{2,0} + + \order(\epsilon) \right) + Z \left( {B_{2,-1}\over \epsilon} + B_{2,0}  + \order(\epsilon) \right) 
+ \left( {C_{2,-2}\over \epsilon^2} + {C_{2,-1}\over \epsilon} + C_{2,0}  + \order(\epsilon) \right)
\nl
{\cal M}_3 &=
Z^3 \left( {A_{3,-2}\over \epsilon^2} + {A_{3,-1}\over \epsilon}  + \order(\epsilon^0) \right)
+ Z^2 \left( {B_{3,-2}\over \epsilon^2} + {B_{3,-1}\over \epsilon}  + \order(\epsilon^0) \right) 
+ Z \left( {C_{3,-2}\over \epsilon^2} + {C_{3,-1}\over \epsilon}  + \order(\epsilon^0) \right) 
\nl
&\quad
+ \left( {D_{3,-3}\over \epsilon^3} + {D_{3,-2}\over \epsilon^2}
 + {D_{3,-1}\over \epsilon}  + \order(\epsilon^0) \right) \,. 
\end{align}
Writing the bare operator as
\begin{align}
O_{\rm bare} = Z_{\cal O} O_{\rm ren}(\mu) \,,
\end{align}
we define the operator renormalization constant $Z_{\cal O}$ in the $\overline{\rm MS}$ scheme, writing 
\begin{align}
Z_{\cal O}&= 1 + {\bar{\alpha}\over 4\pi} {Z_1^{(01)}\over \epsilon} 
+ \left({\bar{\alpha} \over 4\pi}\right)^2 \Bigg[Z^2 {Z_1^{(20)}\over\epsilon}
+ Z {Z_1^{(11)}\over\epsilon} 
+ {Z_2^{(02)}\over\epsilon^2} + {Z_1^{(02)}\over\epsilon}
\Bigg]
\nl
&\quad
+ \left({\bar{\alpha} \over 4\pi}\right)^3 \Bigg[
Z^2 \left( {Z_2^{(21)}\over\epsilon^2} + {Z_1^{(21)}\over \epsilon} \right)
+ \dots 
\Bigg] + \dots \,.
\end{align}
The $\overline{\rm MS}$ coupling, $\bar{\alpha}(\mu)$, is given by 
\begin{align}
{e_0^2 \over 4\pi}(4\pi)^\epsilon \Gamma(1+\epsilon) = \mu^{2\epsilon} Z_\alpha \bar{\alpha}(\mu) \,,
\end{align}
where (for one dynamical electron flavor) 
\begin{align}\label{eq:MSalpha}
Z_\alpha &= 1 + {\bar{\alpha} \over 4\pi} \left( {z_1^{(1)}\over \epsilon} \right)
+ \left(\bar{\alpha} \over 4\pi\right)^2 \left( {z_2^{(2)}\over \epsilon^2}
+ {z_1^{(2)}\over \epsilon} 
\right) + \dots \,,
\end{align} 
with $z_1^{(1)}=4/3$, $z_2^{(2)}=16/9$, $z_1^{(2)}=2$.
The onshell wavefunction renormalization constant for the heavy proton is given by
\begin{align}\label{eq:Zh}
Z_h &= 1 + {\bar{\alpha} \over 4\pi} \left(\lambda\over\mu\right)^{-2\epsilon} 
\left( {Z_{h1}^{(1)} \over \epsilon}  + Z_{h0}^{(1)} + \order(\epsilon) \right)
+ \left({\bar{\alpha} \over 4\pi}\right)^2 \left(\lambda\over\mu\right)^{-4\epsilon} 
\left({Z_{h2}^{(2)} \over \epsilon^2}  + {Z_{h1}^{(2)}\over\epsilon} 
+ \order(\epsilon^0) \right) \,. 
\end{align}
The onshell wavefunction renormalization constant for the 
massless electron field is given by 
\begin{align}\label{eq:Zpsi}
Z_\psi &= 1 + {\bar{\alpha} \over 4\pi} \left(\lambda\over\mu\right)^{-2\epsilon} 
\left( {Z_{\psi 1}^{(1)} \over \epsilon}  + Z_{\psi 0}^{(1)} +  \order(\epsilon) \right)
+ \left({\bar{\alpha} \over 4\pi}\right)^2 \left(\lambda\over\mu\right)^{-4\epsilon} 
\left({Z_{\psi 2}^{(2)} \over \epsilon^2}  + {Z_{\psi 1}^{(2)}\over\epsilon} 
+ \order(\epsilon^0) \right) \,. 
\end{align}
We will require the explicit expressions in arbitrary covariant gauge
$Z_{h 1}^{(1)} = 3-\xi$,  and 
 $Z_{\psi 1}^{(1)} = -\xi$.
The operator renormalization constant $Z_{\cal O}$ is determined by enforcing that the physical amplitude is finite, 
\begin{align}
Z_{\cal O}^{-1} Z_q^\frac12 Z_h^\frac12 {\cal M} = {\rm finite} \,.
\end{align}
After expressing bare coupling in terms of renormalized coupling, we 
find constraints order by order in $\bar{\alpha}$, and order by order in $1/\epsilon$. 
At one-loop order, 
\begin{align}
\order(\alpha/\epsilon):& \quad 
Z_1^{(01)} = B_{1,-1} + \frac12\left( Z_{h1}^{(1)} + Z_{\psi 1}^{(1)} \right) \,.
\end{align} 
At two-loop order, 
\begin{align}
\order(\alpha^2/\epsilon^2):& \quad
Z_2^{(02)} = C_{2,-2} + \left( z_1^{(1)}+ \frac12(Z_{h1}^{(1)}+Z_{\psi 1}^{(1)}) \right) B_{1,-1} 
+ \frac12 (Z_{h2}^{(2)}+Z_{\psi 2}^{(2)}) 
- \frac18 (Z_{h1}^{(1)}-Z_{\psi 1}^{(1)})^2 \,.
\nl
\order(\alpha^2/\epsilon):& \quad
Z_1^{(20)} = A_{2,-1}  \,,
\nl
& \quad
Z_1^{(11)} = B_{2,-1} + (z_1^{(1)}- B_{1,-1}) A_{1,0}  \,,
\nl
& \quad 
Z_1^{(02)} = C_{2,-1} + \frac12 (Z_{h1}^{(2)}+Z_{\psi 1}^{(2)})
+ (z_1^{(1)} - B_{1,-1}) B_{1,0} 
-\frac12 ( Z_{h1}^{(1)} Z_{h0}^{(1)} + Z_{\psi 1}^{(1)} Z_{\psi 0}^{(1)} )
\,.
\nl
\order((\alpha^2/\epsilon)\log\lambda):
& \quad 
C_{2,-2} = \frac14\left( (Z_{h1}^{(1)})^2 + (Z_{\psi 1}^{(1)})^2 \right)
-\frac12 \left( Z_{h2}^{(2)} + Z_{\psi 2}^{(2)} \right)
-\frac12 (z_1^{(1)} - B_{1,-1})B_{1,-1} \,.
\end{align}
The final constraint results from the vanishing of the coefficient 
of $(\alpha^2/\epsilon)\log\lambda$. 
Finally, at three-loop order, we focus on the coefficients involving $Z^2$, 
\begin{align}\label{eq:threeloopdiv}
\order(\alpha^3/\epsilon^2):& \quad
Z_2^{(21)} = B_{3,-2} + \frac12 ( Z_{h1}^{(1)} + Z_{\psi 1}^{(1)} ) A_{2,-1}
+ 2 z_1^{(1)} A_{2,-1} \,.
\nl
\order(\alpha^3/\epsilon^2):
&\quad
Z_1^{(21)} = B_{3,-1} - (z_1^{(1)} - B_{1,-1}) (A_{1,0})^2 - A_{1,0}B_{2,-1}
+ (2z_1^{(1)}-B_{1,-1})A_{2,0} - B_{1,0}A_{2,-1} \,.
\nl
\order((Z^2\alpha^3/\epsilon)\log\lambda):
&\quad
B_{3,-2} = A_{2,-1}\left(-\frac43 z_1^{(1)} + B_{1,-1}\right) \,.
\end{align}
As usual, the operator anomalous dimension is determined by the coefficient of 
$1/\epsilon$ in $Z_{\cal O}$: 
\begin{align}\label{eq:gammafromZ}
{d\over d\log\mu} O_{\rm ren} = 
\gamma_{\cal O} O_{\rm ren} \,,
\quad
\gamma_{\cal O} &= -2\alpha {\partial \over \partial \alpha} [Z_{\cal O}]_1 \,.
\end{align}
In particular, our focus is on $\gamma_2^{(1)} = -6 Z_1^{(21)}$, where $Z_1^{(21)}$ is determined using  Eq.~(\ref{eq:threeloopdiv}). 
Through three-loop order, 
contributions involving at most a single power of $\alpha$ not accompanied by $Z$ 
involve: the $A_{i,j}$ and $B_{i,j}$ coefficients in 
the amputated amplitudes from (\ref{eq:amp}); 
the coefficient $z_1^{(1)}$ in $Z_\alpha$ 
given after Eq.~(\ref{eq:MSalpha});
and the wavefunction renormalization factors 
$Z_{h1}^{(1)}$ and $Z_\psi^{(1)}$
given after Eqs.~(\ref{eq:Zh}) and (\ref{eq:Zpsi}).

\section{Amplitude evaluation \label{sec:eval}}

We proceed to evaluate amplitudes at one, two and three loops (see the Appendices for details). 

\subsection{One-loop amplitude}

Consider the amputated diagrams at one-loop. 
The contribution of order $Z^1$ vanishes due to rotational symmetry:
\be
\parbox{30mm}{
\begin{fmfgraph*}(80,40)
  \fmfleftn{l}{3}
  \fmfrightn{r}{3}
  \fmfbottomn{b}{5}
  \fmf{phantom}{l2,v,r2}
  \fmffreeze
  \fmf{fermion}{r3,x,v}
  \fmf{double}{l2,v}
  \fmffreeze
  \fmf{photon}{x,b4}
  \fmfv{decor.shape=cross}{b4}
    \fmfv{decor.shape=circle,decor.filled=shaded}{v}
\end{fmfgraph*}
}
= 0 \,,
\ee
implying 
\begin{align}
A_{1,0}=0\,,
\qquad 
A_{1,1}&=0 \,.
\end{align}
Next for the vertex correction of order $Z^0$, 
\vspace{6pt}
\be
\parbox{30mm}{
\begin{fmfgraph*}(100,65)
  \fmfleftn{l}{3}
  \fmfrightn{r}{3}
  \fmfbottomn{b}{5}
  \fmf{phantom}{l2,v,r2}
  \fmffreeze
  \fmf{fermion}{r3,x,v}
  \fmf{double}{l2,y,v}
  \fmffreeze
  \fmf{photon,right}{x,y}
\fmfv{decor.shape=circle,decor.filled=shaded}{v}
\end{fmfgraph*}
}
= e_0^2\,\xi\, M(1,1,\lambda) 
= {e_0^2\over (4\pi)^2}[(4\pi)^\epsilon \Gamma(1+\epsilon)]\lambda^{-2\epsilon}\xi \left({1\over \epsilon} + 1 + \order(\epsilon^1)\right) 
\,,
\ee
where integral $M(1,1,\lambda)$ is given in Appendix~\ref{sec:elementary}, 
implying 
\begin{align}
B_{1,-1}=\xi\,,
\qquad 
B_{1,0}=\xi \,.
\end{align}

\subsection{Two-loop amplitude}
UV divergences from iterated Coulomb insertions first arise at two loops,
\begin{align}\label{eq:skel}
&\parbox{40mm}{
\begin{fmfgraph*}(100,60)
  \fmfleftn{l}{3}
  \fmfrightn{r}{3}
  \fmfbottomn{b}{7}
  \fmf{phantom}{l2,v,r2}
  \fmf{phantom}{l1,b5,b6,r1}
  \fmffreeze
  \fmf{fermion}{r3,x,y,v}
  \fmf{double}{l2,v}
  \fmffreeze
  \fmf{photon}{x,b6}
  \fmf{photon}{y,b5}
  \fmfv{decor.shape=cross}{b5}
  \fmfv{decor.shape=cross}{b6}
      \fmfv{decor.shape=circle,decor.filled=shaded}{v}
\end{fmfgraph*}
}
\nl
&= 
Z^2 e_0^4 \int (\dd^dq) (\dd^dq^\prime) (2\pi)\delta(q^0) (2\pi) \delta(q^{\prime 0}) 
{1 \over q^2-\lambda^2}
{1\over q^{\prime 2} - \lambda^2}
{ \slash{q} + \slash{q}^\prime \over (q+q^\prime )^2}\gamma^0 {\slash{q}\over q^2}\gamma^0
\nl
&=
Z^2 e_0^4 \left(\frac12\right) \left[ L(1,1,\lambda) A(1,\lambda) + B(1,1,1,\lambda) \right] 
\nl 
&= Z^2 e_0^4 
\left[ {\Gamma(1+\epsilon) \lambda^{-2\epsilon}
\over (4\pi)^{2-\epsilon} }\right]^2 
\left[ {2\pi^2 \over \epsilon} - 4\pi^2 + \order(\epsilon^0) \right]
\,,
\end{align}
where $(\dd^dk)=d^dk/(2\pi)^d$, and the integrals $A$, $B$ and $L$ are given in Appendix~\ref{sec:elementary}. 
This implies (in arbitrary covariant gauge)
\begin{align}
 A_{2,-1}= 2\pi^2 \,,
\qquad
 A_{2,0} =  -4\pi^2 \,.
\end{align}
Here we have used for the numerator, 
\begin{align}
N = (\slash{q}+\slash{q}^\prime)\gamma^0 \slash{q}\gamma^0 
\to \frac14 {\rm Tr}(N) 
= \bm{q}\cdot (\bm{q}+\bm{q}^\prime)
= \frac12\left[ (\bm{q}+\bm{q}^\prime)^2 + (\bm{q}^2+\lambda^2) - (\bm{q}^{\prime 2} + \lambda^2) 
\right] \,.
\end{align}

While they are not needed for the $\order(Z^2\alpha^3)$ anomalous dimension, the nonvanishing contributions at 
$\order(Z\alpha^2)$ are as follows: 
\begin{align}
\nl
&\parbox{30mm}{
\begin{fmfgraph*}(100,60)
  \fmfleftn{l}{3}
  \fmfrightn{r}{3}
  \fmfbottomn{b}{7}
  \fmf{phantom}{l2,v,r2}
  \fmf{phantom}{l1,b5,b6,r1}
  \fmffreeze
  \fmf{fermion}{r3,x,y,v}
  \fmf{double}{l2,z,v}
  \fmffreeze
  \fmf{photon}{x,b6}
  \fmf{photon,right}{y,z}
  \fmfv{decor.shape=cross}{b6}
      \fmfv{decor.shape=circle,decor.filled=shaded}{v}
\end{fmfgraph*}
}
\quad + \quad 
\parbox{30mm}{
\begin{fmfgraph*}(100,60)
  \fmfleftn{l}{3}
  \fmfrightn{r}{3}
  \fmfbottomn{b}{7}
  \fmf{phantom}{l2,v,r2}
  \fmf{phantom}{l1,b5,b6,r1}
  \fmffreeze
  \fmf{fermion}{r3,x,y,v}
  \fmf{double}{l2,z,v}
  \fmffreeze
  \fmf{photon,right}{x,z}
  \fmf{photon}{y,b5}
  \fmfv{decor.shape=cross}{b5}
      \fmfv{decor.shape=circle,decor.filled=shaded}{v}
\end{fmfgraph*}
}
=Ze_0^4 \left(\frac12\right)\left[ 
 L(1,1,\lambda)A(1,\lambda) +
B(1,1,1,\lambda)\right] \,,
\nl
\end{align}
implying (in arbitrary covariant gauge)
\begin{align}
B_{2,-1} = 2\pi^2 \,,
\qquad
B_{2,0} &= -4\pi^2 \,. 
\end{align}

\subsection{Three-loop amplitude}

\begin{figure}[t]
\centering
\begin{subfigure}[]{0.25 \textwidth}
\parbox{40mm}{
\begin{fmfgraph*}(120,60)
    \fmfstraight
  \fmfleftn{l}{3}
  \fmfrightn{r}{3}
  \fmfbottomn{b}{9}
  \fmf{phantom}{l2,v,r2}
  \fmffreeze
  \fmf{fermion}{r3,x,y,z,v}
  \fmf{double}{l2,w,v}
  \fmffreeze
  \fmf{photon}{x,b8}
  \fmf{photon}{y,b7}
  \fmf{photon,right}{z,w}
  \fmfv{decor.shape=cross}{b8}
  \fmfv{decor.shape=cross}{b7}
      \fmfv{decor.shape=circle,decor.filled=shaded}{v}
\end{fmfgraph*}
}
\subcaption{$(a)$}
\end{subfigure}
\begin{subfigure}[]{0.25\textwidth}
\parbox{30mm}{
\begin{fmfgraph*}(120,60)
    \fmfstraight
  \fmfleftn{l}{3}
  \fmfrightn{r}{3}
  \fmfbottomn{b}{9}
  \fmf{phantom}{l2,v,r2}
  \fmffreeze
  \fmf{fermion}{r3,x,y,z,v}
  \fmf{double}{l2,w,v}
  \fmffreeze
  \fmf{photon}{x,b8}
  \fmf{photon}{z,b6}
  \fmf{photon,right}{y,w}
  \fmfv{decor.shape=cross}{b8}
  \fmfv{decor.shape=cross}{b6}
      \fmfv{decor.shape=circle,decor.filled=shaded}{v}
\end{fmfgraph*}
}
\subcaption{$(b)$}
\end{subfigure}
\begin{subfigure}[]{0.25\textwidth}
\parbox{30mm}{
\begin{fmfgraph*}(120,60)
\fmfstraight
  \fmfleftn{l}{3}
  \fmfrightn{r}{3}
  \fmfbottomn{b}{9}
  \fmf{phantom}{l2,v,r2}
  \fmf{phantom}{l1,b6,b7,b8,r1}
  \fmffreeze
  \fmf{fermion}{r3,x,y,z,v}
  \fmf{double}{l2,w,v}
  \fmffreeze
  \fmf{photon}{z,b6}
  \fmf{photon}{y,b7}
  \fmf{photon,right}{x,w}
  \fmfv{decor.shape=cross}{b6}
  \fmfv{decor.shape=cross}{b7}
      \fmfv{decor.shape=circle,decor.filled=shaded}{v}
\end{fmfgraph*}
}\subcaption{$(c)$}
\end{subfigure}
\begin{subfigure}[]{0.25\textwidth}
\vspace{5mm}
\parbox{40mm}{
\begin{fmfgraph*}(120,60)
    \fmfstraight
  \fmfleftn{l}{3}
  \fmfrightn{r}{3}
  \fmfbottomn{b}{16}
  \fmf{phantom}{l2,v1,v2,r2}
  \fmffreeze
  \fmf{fermion}{r3,x,y,z,w,v1}
  \fmf{double}{l2,v1}
  \fmffreeze
  \fmf{photon}{x,b14}
  \fmf{photon}{y,b12}
  \fmf{photon,right=1.5}{z,w}
  \fmfv{decor.shape=cross}{b14}
  \fmfv{decor.shape=cross}{b12}
      \fmfv{decor.shape=circle,decor.filled=shaded}{v1}
\end{fmfgraph*}
}
\subcaption{$(p1)$}
\end{subfigure}
\begin{subfigure}[]{0.25\textwidth}
\vspace{5mm}
\parbox{40mm}{
\begin{fmfgraph*}(120,60)
\fmfstraight
  \fmfleftn{l}{3}
  \fmfrightn{r}{3}
  \fmfbottomn{b}{16}
  \fmf{phantom}{l2,v1,v2,r2}
  \fmffreeze
  \fmf{fermion}{r3,x,y,z,w,v1}
  \fmf{double}{l2,v1}
  \fmffreeze
  \fmf{photon}{x,b14}
  \fmf{photon}{w,b8}
  \fmf{photon,right=1.5}{y,z}
  \fmfv{decor.shape=cross}{b14}
  \fmfv{decor.shape=cross}{b8}
      \fmfv{decor.shape=circle,decor.filled=shaded}{v1}
\end{fmfgraph*}
}
\subcaption{$(p2)$}
\end{subfigure}
\begin{subfigure}[]{0.25\textwidth}
\vspace{5mm}
\parbox{40mm}{
\begin{fmfgraph*}(120,60)
\fmfstraight
  \fmfleftn{l}{3}
  \fmfrightn{r}{3}
  \fmfbottomn{b}{16}
  \fmf{phantom}{l2,v1,v2,r2}
  \fmffreeze
  \fmf{fermion}{r3,x,y,z,w,v1}
  \fmf{double}{l2,v1}
  \fmffreeze
  \fmf{photon}{x,b14}
  \fmf{photon}{z,b10}
  \fmf{photon,right}{y,w}
  \fmfv{decor.shape=cross}{b14}
  \fmfv{decor.shape=cross}{b10}
      \fmfv{decor.shape=circle,decor.filled=shaded}{v1}
\end{fmfgraph*}
}
\subcaption{$(v1)$}
\end{subfigure}
\begin{subfigure}[]{0.25\textwidth}
\vspace{5mm}
\parbox{40mm}{
\begin{fmfgraph*}(120,60)
\fmfstraight
  \fmfleftn{l}{3}
  \fmfrightn{r}{3}
  \fmfbottomn{b}{16}
  \fmf{phantom}{l2,v1,v2,r2}
  \fmffreeze
  \fmf{fermion}{r3,x,y,z,w,v1}
  \fmf{double}{l2,v1}
  \fmffreeze
  \fmf{photon}{y,b12}
  \fmf{photon}{w,b8}
  \fmf{photon,right}{x,z}
  \fmfv{decor.shape=cross}{b12}
  \fmfv{decor.shape=cross}{b8}
      \fmfv{decor.shape=circle,decor.filled=shaded}{v1}
\end{fmfgraph*}
}
\subcaption{$(v2)$}
\end{subfigure}
\begin{subfigure}[]{0.25\textwidth}
\vspace{5mm}
\parbox{40mm}{
\begin{fmfgraph*}(120,80)
  \fmfleftn{l}{3}
  \fmfrightn{r}{3}
  \fmfbottomn{b}{10}
  \fmf{phantom}{l2,v1,v2,r2}
  \fmffreeze
  \fmf{fermion}{r3,x,y,v1}
  \fmf{double}{l2,v1}
  \fmffreeze
  \fmf{photon}{x,b8}
  \fmf{photon}{y,a1}
  \fmf{photon}{a2,b6}
  \fmf{fermion,left,tension=0.5}{a1,a2,a1}
  \fmfv{decor.shape=cross}{b8}
  \fmfv{decor.shape=cross}{b6}
      \fmfv{decor.shape=circle,decor.filled=shaded}{v1}
\end{fmfgraph*}
}
\subcaption{$(b1)$}
\end{subfigure}
\begin{subfigure}[]{0.25\textwidth}
\vspace{5mm}
\parbox{40mm}{
\begin{fmfgraph*}(120,80)
  \fmfleftn{l}{3}
  \fmfrightn{r}{3}
  \fmfbottomn{b}{10}
  \fmf{phantom}{l2,v1,v2,r2}
  \fmffreeze
  \fmf{fermion}{r3,x,y,v1}
  \fmf{double}{l2,v1}
  \fmffreeze
  \fmf{photon}{y,b6}
  \fmf{photon}{x,a1}
  \fmf{photon}{a2,b8}
  \fmf{fermion,left,tension=0.5}{a1,a2,a1}
  \fmfv{decor.shape=cross}{b8}
  \fmfv{decor.shape=cross}{b6}
      \fmfv{decor.shape=circle,decor.filled=shaded}{v1}
\end{fmfgraph*}
}
\subcaption{$(b2)$}
\end{subfigure}

\vspace{24pt}

\begin{subfigure}[]{0.25\textwidth}
\vspace{5mm}
\parbox{40mm}{
\begin{fmfgraph*}(140,60)
    \fmfstraight
  \fmfleftn{l}{3}
  \fmfrightn{r}{3}
  \fmfbottomn{b}{16}
  \fmf{phantom}{l2,v1,v2,r2}
  \fmffreeze
  \fmf{fermion}{r3,x,y,z,w,v1}
  \fmf{double}{l2,v1}
  \fmffreeze
  \fmf{photon,label=$\vb{q}\hspace{-18pt}$}{y,b12}
  \fmf{photon,label=$\vb{q}'$}{z,b10}
  \fmf{photon,right,label=$k$}{x,w}
  \fmfv{decor.shape=cross}{b12}
  \fmfv{decor.shape=cross}{b10}
\fmfv{decor.shape=circle,decor.filled=shaded}{v1}
\end{fmfgraph*}
}
\subcaption{$(w)$}
\end{subfigure}
    \caption{Three-loop amputated amplitude at order $Z^2\alpha^3$. Our momentum routing is illustrated with $\mathcal{M}^w$ where $\vb{q}'=\vb{p}-\vb{q}$. The photon labeled with $k$ has a non-zero $k_0$ component whereas those labeled with three-vectors do not. \label{fig:threeloopdiagrams}}
\end{figure}

We focus on the contributions at order $Z^2\alpha^3$.  
Let us define the following integrals, 
\begin{align}\label{eq:Iintegrals}
&I^{(b)}(a_1,a_2,a_3,a_4,a_5,a_6) 
\nl
&= \int(\dd\omega)(\dd k)(\dd q)(\dd p)\,\omega^b\,
{1\over [\omega^2+(\bm{k}+\bm{p})^2]^{a_1}}{1\over (\vb{p}^2)^{a_2}}{1\over [(\bm{p}-\bm{q})^2]^{a_3}}{1\over [\omega^2+(\bm{k}+\bm{q})^2]^{a_4}}{1\over (\vb{q}^2)^{a_5}}{1\over \vb{q}^2+\lambda^2}
\times \nl &\qquad \times 
{1\over (\omega^2+\bm{k}^2)^{a_6}}{1\over \omega^2+\bm{k}^2+\lambda^2} \,,
\end{align}
where $(\dd \omega)=d\omega/(2\pi)$, and $(\dd k)=d^Dk/(2\pi)^D$.  
Integrals written without a superscript imply $b=0$: $I\equiv I^{(0)}$.  
In the following we illustrate the procedure by evaluating diagrams $(a)$, $(b)$, $(c)$, $(b1)$, and $(b2)$ in Feynman gauge.  Details of the evaluation for all diagrams in arbitrary covariant gauge are presented in the Appendix. 

\subsubsection{Jaus diagrams}

Consider first diagrams $(a)$, $(b)$, $(c)$, considered in Ref.~\cite{Jaus:1972hua}.  For the first of these, in Feynman gauge ($\xi=1$), 
\begin{multline}\label{eq:Ma}
{\cal M}^a\big|_{\xi=1} =
-Z^2 e^6 \int(\dd \omega)(\dd k)(\dd q)(\dd q^\prime)
{1\over -k^0+i0}{1\over \bm{q}^2+\lambda^2}{1\over \bm{q}^{\prime 2}+\lambda^2}{1\over \omega^2+\bm{k}^2+\lambda^2}
{1\over \omega^2+(\bm{q}+\bm{q}^\prime+\bm{k})^2} \times
\\
\times
{1\over (\bm{q}+\bm{q}^\prime)^2}
{1\over \bm{q}^2} 
N^a
\end{multline}
where $k^0 \equiv ik_E^0 \equiv i\omega$ and the numerator is
\begin{align}
N^a &= (\slash{q}+\slash{q}^\prime+\slash{k})\gamma^0(\slash{q}+\slash{q}^\prime)\gamma^0 \slash{q} \gamma^0 
\to \frac14 {\rm Tr}(N^a) 
= k^0 \bm{q}\cdot (\bm{q}+\bm{q}^\prime) \,.
\end{align}
Note that the numerator evaluation is valid in $d=4-2\epsilon$ dimensions.
With the substitution 
\begin{align}
\bm{q}+\bm{q}^\prime = \bm{p} \,,
\end{align}
the integral becomes 
\begin{align}\label{eq:Maexpand}
{\cal M}^a \big|_{\xi=1}
&= Z^2 e^6 \left[ \frac12\left(-\bm{3}^-+\bm{2}^-+\bm{5}^-\right)I_{111010} + \delta I^a \right] \,,
\end{align}
where we introduce the notation 
$\bm{n}^\pm I_{a_1 \cdots a_{n-1} \, a_n \, a_{n+1}\cdots a_6} = 
I_{a_1 \cdots a_{n-1} (a_n\pm 1) a_{n+1} \cdots a_6}$.  
The term $\delta I^a$ results from writing 
\begin{align}\label{eq:deltareplace}
{1\over (\bm{p}-\bm{q})^2+\lambda^2} 
= {1\over (\bm{p}-\bm{q})^2} 
- {\lambda^2 \over (\bm{p}-\bm{q})^2 (\bm{p}-\bm{q})^2+\lambda^2} \,. 
\end{align}
After simplification we have
\begin{align}
\delta I^a &= -{\lambda^2\over 2}
\int(\dd \omega)(\dd k)(\dd q)(\dd p)
{1\over \omega^2+\bm{k}^2+\lambda^2}{1\over \omega^2+(\bm{k}+\bm{p})^2}
\left[ -(\bm{p}-\bm{q})^2 + \bm{q}^2 + \bm{p}^2 \right]
\times
\nl &\quad \times
{1 \over (\bm{p}-\bm{q})^2[(\bm{p}-\bm{q})^2+\lambda^2]}
{1 \over \bm{q}^2 (\bm{q}^2+\lambda^2)}{1\over \bm{p}^2} \,.
\end{align}
The integral is finite apart from the subdivergence at $\omega, k \to \infty$.  Up to finite terms (we use ``$\sim$" to denote equality up to finite terms as $\epsilon\to 0$), we thus have
\begin{align}\label{eq:deltaIa}
\delta I^a &\sim 
-{\lambda^2\over 2} 
\bigg[\int(\dd \omega)(\dd k)
{1\over \omega^2+\bm{k}^2+\lambda^2}{1\over \omega^2+\bm{k}^2}\bigg]
\bigg[\int (\dd q)(\dd p)
{1 \over (\bm{p}-\bm{q})^2[(\bm{p}-\bm{q})^2+\lambda^2]}
{1 \over \bm{q}^2 (\bm{q}^2+\lambda^2)}\bigg]
\nl
&\sim -{\lambda^2\over 2} M(1,1,\lambda) 
\bigg[ \lim_{r\to 0}\left({1\over\lambda^2}(f_1(0,r)-f_1(\lambda,r))\right)\bigg]^2 
\nl
&=
\left[(4\pi)^\epsilon \Gamma(1+\epsilon) \over (4\pi)^2\right]^3 
\left( -{8\pi^2 \over \epsilon} + \order(\epsilon^0)\right)
\,,
\end{align}
where the $(\dd q)(\dd p)$ integration may be performed in $D=3$. 
Evaluating integrals, 
\begin{align}
{\cal M}^a\big|_{\xi=1} &= Z^2 e^6  \left[(4\pi)^\epsilon \Gamma(1+\epsilon) \over (4\pi)^2\right]^3 \pi^2 \left( {4\over 3\epsilon^2} - {4\over \epsilon} + \order(\epsilon^0) \right) \,.
\end{align}
Similar manipulations for diagram~($b$) yield 
\begin{align}
{\cal M}^b\big|_{\xi=1} &= Z^2 e^6 
\left[ \left(-\frac12\bm{3}^- +\frac12 \bm{2}^- +\frac12\bm{5}^- +\bm{4}^- -\bm{6}^- \right)I_{101110} + \delta I^b\right] 
\nl
&= Z^2 e^6  \left[(4\pi)^\epsilon \Gamma(1+\epsilon) \over (4\pi)^2\right]^3 \pi^2 \left[{1\over \epsilon}\left(-4 + {8\pi^2 \over 9} \right) + \order(\epsilon^0) \right] \,.
\end{align}
In this case, the second term on the right-hand side of Eq.~(\ref{eq:deltareplace}) leads to a finite integral, $\delta I^b \sim 0$. 
Similarly for diagram~($c$), 
\begin{align}
{\cal M}^c\big|_{\xi=1} &= Z^2 e^6 \bigg[ -4 I^{(2)}_{101101} + \left(\bm{1}^- + \bm{4}^- + \bm{6}^- -\frac12 \bm{2}^- -\frac12\bm{3}^- - \frac12 \bm{5}^-\right)I_{101101} + \delta I^c \bigg]
\nl
&= Z^2 e^6  \left[(4\pi)^\epsilon \Gamma(1+\epsilon) \over (4\pi)^2\right]^3 \pi^2 \left[{2\over 3\epsilon^2} + 
{1 \over \epsilon}\left( -2 + {8\pi^2 \over 9} \right) + \order(\epsilon^0) \right] \,,
\end{align}
where $\delta I^c \sim 0$. 

We are unable to make a direct comparison to the results of Ref.~\cite{Jaus:1972hua} ({\it cf}. also Ref.~\cite{Jaus:1986te}), which considered only diagrams $(a)$, $(b)$ and $(c)$ in Fig.~\ref{fig:threeloopdiagrams}.  
We note that UV subdivergences appear in these diagrams that are not regulated by working at 
finite proton mass.\!\footnote{
The proton propagator used in Ref.~\cite{Jaus:1972hua}
corresponds to the replacement 
$i/(k^0+i0) \to 2M i/(k^2 +2M k^0+i0)$   
in the heavy particle propagator, {\it cf}. Fig.~\ref{fig:prop}.
}
For example, inspection of diagram $(c)$, cf. Eq.~(\ref{app_eq:Mc}) below, 
shows an unregulated divergence when $|\bm{p}|,\,|\bm{q}| \to \infty$ at fixed $\omega,\,|\bm{k}|$.\!\footnote{In Eq.~(3) of Ref.~\cite{Jaus:1972hua} a nuclear form factor is included, which would regulate this subdivergence; however this form factor is later set to unity, as discussed beneath Eq.~(4) of Ref.~\cite{Jaus:1972hua}, and can therefore not remove the subdivergence.
}

\subsubsection{Vacuum polarization diagrams \label{bubble-eval} }
Diagram~($b2$) is obtained by inserting 
\begin{align}\label{eq:b1}
{q^2 \over q^2 -\lambda^2} \Pi(q^2) 
= {e^2 \over (4\pi)^2}[(4\pi)^\epsilon\Gamma(1+\epsilon)] (\bm{q}^2)^{-\epsilon}
{\bm{q}^2\over \bm{q}^2+\lambda^2} \left[ -{8\over \epsilon}\beta(2-\epsilon,2-\epsilon) \right]~, 
\end{align}
into the skeleton integral (\ref{eq:skel}).  A straightforward integration yields the result
\begin{align}\label{eq:b2}
{\cal M}^{b2}\big|_{\xi=1} &= Z^2 e^6  \left[(4\pi)^\epsilon \Gamma(1+\epsilon) \over (4\pi)^2\right]^3 \pi^2 \left[-{16\over 9\epsilon^2} + {1 \over \epsilon}\left({32\over 3} \log(2) - {32\over 27} \right) + \order(\epsilon^0) \right] \,.
\end{align}
Similarly, inserting $[q^{\prime 2}/(q^{\prime 2}-\lambda^2)]\Pi(q^{\prime 2})$ from Eq.~(\ref{eq:b1}) 
into the skeleton integral (\ref{eq:skel}), we obtain the result 
\begin{align}
{\cal M}^{b1}\big|_{\xi=1} &= Z^2 e^6  \left[(4\pi)^\epsilon \Gamma(1+\epsilon) \over (4\pi)^2\right]^3 \pi^2 \left[-{16\over 9\epsilon^2} + {1 \over \epsilon}\left(-{32\over 3} \log(2) + {32\over 27} \right) + \order(\epsilon^0) \right] \,.
\end{align}

\subsubsection{Summary of three-loop diagrams}

As described in the Appendix, 
the remaining diagrams are obtained by similar manipulations to those described for diagrams $(a)$, $(b)$, and $(c)$: expanding on the integral basis, extracting subdivergences and writing remainders as $1/\epsilon$ times convolution integrals in integer dimension.  
Collecting results for all diagrams, including gauge dependent terms, the results are 
\begin{align} \label{eq:Msummary}
{\cal M}^a &= Z^2 e^6  \left[(4\pi)^\epsilon \Gamma(1+\epsilon) \over (4\pi)^2\right]^3 \pi^2 \left[ {4\over 3\epsilon^2} - {4\over \epsilon} + 
(1-\xi)\left(-{2\over \epsilon^2}+{10\over 3\epsilon} \right) 
+ \order(\epsilon^0) \right] \,,
\nl
{\cal M}^b &= Z^2 e^6  \left[(4\pi)^\epsilon \Gamma(1+\epsilon) \over (4\pi)^2\right]^3 \pi^2 \left[{1\over \epsilon}\left(-4 + {8\pi^2 \over 9} \right) 
+ (1-\xi)\left({2\over 3\epsilon^2}+{2\over 3\epsilon} \right) 
+ \order(\epsilon^0) \right] \,,
\nl
{\cal M}^c &= Z^2 e^6  \left[(4\pi)^\epsilon \Gamma(1+\epsilon) \over (4\pi)^2\right]^3 \pi^2 \left[{2\over 3\epsilon^2} + 
{1 \over \epsilon}\left( -2 + {8\pi^2 \over 9} \right) + 
 (1-\xi)\left(-{2\over 3\epsilon^2}-{2\over \epsilon} \right) 
+ \order(\epsilon^0) \right] \,,
\nl
{\cal M}^{b1} &= Z^2 e^6  \left[(4\pi)^\epsilon \Gamma(1+\epsilon) \over (4\pi)^2\right]^3 \pi^2 \left[-{16\over 9\epsilon^2} + {1 \over \epsilon}\left(-{32\over 3} \log(2) + {32\over 27} \right) + \order(\epsilon^0) \right] \,,
\nl
{\cal M}^{b2} &= Z^2 e^6  \left[(4\pi)^\epsilon \Gamma(1+\epsilon) \over (4\pi)^2\right]^3 \pi^2 \left[-{16\over 9\epsilon^2} + {1 \over \epsilon}\left({32\over 3} \log(2) - {32\over 27} \right) + \order(\epsilon^0) \right] \,,
\nl
{\cal M}^{p1} &= Z^2 e^6  \left[(4\pi)^\epsilon \Gamma(1+\epsilon) \over (4\pi)^2\right]^3 \pi^2 \left[ 
-{4\over 3\epsilon^2} + {16\over 3\epsilon} 
+ (1-\xi)\left(-{4\over 3\epsilon} \right)
+ \order(\epsilon^0) \right] \,,
\\
{\cal M}^{p2} &= Z^2 e^6  \left[(4\pi)^\epsilon \Gamma(1+\epsilon) \over (4\pi)^2\right]^3 \pi^2 \left[ 
-{4\over 3\epsilon^2} + {4\over \epsilon}
+
(1-\xi)\left( {10\over 3\epsilon^2} + {10\over \epsilon}\right)
+ \order(\epsilon^0) \right] \,,
\nl
{\cal M}^{v1} &= Z^2 e^6  \left[(4\pi)^\epsilon \Gamma(1+\epsilon) \over (4\pi)^2\right]^3 \pi^2 \left[ 
{4\over 3\epsilon^2}
- \frac{8}{\epsilon} 
+(1-\xi)\left(-{2\over\epsilon^2} - {14\over 3\epsilon} \right)
+ \order(\epsilon^0) \right] \,,
\nl
{\cal M}^{v2} &= Z^2 e^6  \left[(4\pi)^\epsilon \Gamma(1+\epsilon) \over (4\pi)^2\right]^3 \pi^2 \left[ 
{4\over 3\epsilon^2} - {20\over 3\epsilon}
+
(1-\xi)\left(-{4\over 3\epsilon^2} - {4\over \epsilon}\right)
+ \order(\epsilon^0) \right] \,,
\nl
{\cal M}^{w} &= Z^2 e^6  \left[(4\pi)^\epsilon \Gamma(1+\epsilon) \over (4\pi)^2\right]^3 \pi^2 \left[ 
{1\over \epsilon} \left( 8 - {8\pi^2 \over 9} \right) 
+ \order(\epsilon^0) \right] \,. \nonumber
\end{align}
Using the above results we obtain 
\begin{equation}
    B_{3,-2} = -{14 \pi^2 \over 9} -2\pi^2 (1-\xi) \,, \qquad
    B_{3,-1}= \pi ^2 \left(\frac{8 \pi ^2}{9}-\frac{22}{3}\right)+2 \pi ^2 (1-\xi )~.
\end{equation}

\section{Anomalous dimension \label{sec:anom}}

Using Eq.~(\ref{eq:gammafromZ}), 
the expansion of the anomalous dimension
reads
\begin{align}
\gamma_{\cal O} &= 
-2 {\bar{\alpha} \over 4\pi} Z_1^{(01)} 
-4 \left(\bar{\alpha} \over 4\pi\right)^2 
\left[ Z^2 Z_1^{(20)} + Z Z_1^{(11)} + Z_1^{(02)} \right]
- 6 \left(\bar{\alpha} \over 4\pi\right)^3 
\left[ Z^2 Z_1^{(21)} + 
Z Z_1^{(11)} + Z_1^{(02)} \right] + \dots
\end{align}
Using the results above, we may read off 
the results at one and two-loop order, 
\begin{align} 
\gamma_0^{(1)} &= -2 Z_1^{(01)} = -3 \,,
\nl
\gamma_1^{(0)} &= -4 Z_1^{(20)} = -8\pi^2 \,,
\nl
\gamma_1^{(1)} &= -4 Z_1^{(11)} = -8\pi^2 \,.
\end{align}
In particular, $\gamma_1^{(0)}$ and $\gamma_1^{(1)}$ obey the relation 
(\ref{eq:evenloopsymm}). 
At three-loop order we verify the consistency condition for $B_{3,-2}$, Eq.~(\ref{eq:threeloopdiv}). For the anomalous dimension at $\order(Z^2\alpha^3)$ we  obtain 
\begin{align}
\gamma_2^{(1)} &= -6 Z_1^{(21)} = -6\left( B_{3,-1} - {26\pi^2 \over 3} - 2\pi^2(1-\xi) \right) 
= 16\pi^2 \left( 6 - {\pi^2\over 3} \right) \,.
\end{align}
This is our main result, and represents new perturbative input for the evaluation of long-distance QED radiative corrections to nuclear beta decay. 

Using the symmetries of Section~\ref{sec:symm}, and existing results for heavy-light currents~\cite{Sirlin:1967zza,Ji:1991pr,Chetyrkin:2003vi}, we obtain the anomalous dimension of the heavy-heavy-light operators complete through three-loop order. Moreover, fixing the leading-$Z$ contribution at four loops with the Dirac equation \cite{Hill:2023bfh} and again using the symmetries of Section~\ref{sec:symm} we also fix the $Z^3\alpha^4$ coefficient of the anomalous dimension. Our results are summarized in Table~\ref{tab:summary}, where one sees the remaining unknown coefficient at four-loop order marked with ``\qMark''.
\begin{table}[h]
\centering
\begin{tabular}{|c||*{4}{c|}}\hline\hline
\backslashbox{$Z^n$}{Loops}
&\makebox[3em]{1-loop}&\makebox[3em]{2-loop}&\makebox[3em]{3-loop}
&\makebox[3em]{4-loop}\\\hline\hline
& & & & \\[-9pt]
$Z^0$ & $~~~\gamma_0^{(1)}=-3~~~$ & $\gamma_1^{(2)}= -16 \zeta_2 + \frac52 + \frac{10}{3} n_e$ & $\gamma_2^{(3)} = (\text{see~caption})$ & $\gamma_3^{(4)}=(\text{see~caption})$ \\[6pt] 
$Z^1$ & $~~\gamma_{0}^{(0)} = 0~~~~~$ &  
$\gamma_{1}^{(1)} =\gamma_{0}^{(1)}~~$ 
& $\gamma_{2}^{(2)} = \gamma_2^{(1)}$ & $\gamma_3^{(3)}=\gamma_3^{(2)}-\gamma_3^{(0)}$\\[6pt]
$Z^2$ & -- & $\gamma_{1}^{(0)} = -8\pi^2$ & $\gamma_{2}^{(1)} = 16\pi^2\left(6-\frac{\pi^2}{3}\right)$ & $\gamma_3^{(2)}=~$\qMark \\[6pt]
$Z^3$ & -- & -- & $\gamma_{2}^{(0)} = 0$~~~ & $\gamma_{3}^{(1)}=2\gamma_{3}^{(0)}$ \\[6pt]
$Z^4$ & -- & -- & --& $\gamma_{3}^{(0)} =-32\pi^4 $\\[3pt]\hline\hline
\end{tabular}
\caption{Summary of known coefficients in the perturbative expansion of $\gamma_{\mathcal{O}}$ for heavy-heavy-light operators for a $U(1)$ gauge theory with $n_e$ light fermions. 
The coefficients at $O(Z^n\alpha^n)$ are fixed by solutions of the Dirac equation and vanish for odd-loop orders due to rotational invariance \cite{Hill:2023bfh}. The one-loop result for $\gamma_0^{(1)}$ was first computed by Sirlin \cite{Sirlin:1967zza} and the two-loop result for $\gamma_{1}^{(2)}$ by Ji and Ramsey-Musolf \cite{Ji:1991pr}. The three-loop result for $\gamma_{2}^{(3)}$ is given by $\gamma_2^{(3)}=-80 \zeta_4 -36 \zeta_3 + 64\zeta_2 -\frac{37}{2} + n_e\left( -\frac{176}{3}\zeta_3 + \frac{448}{9}\zeta_2 + \frac{470}{9} \right) + \frac{140}{27} n_e^2$ as computed by Chetyrkin and Grozin \cite{Chetyrkin:2003vi}. The four-loop result for $\gamma_3^{(4)}$ has recently been computed by Grozin and is presented in analytic form in Eq.~(3.3) of Ref.~\cite{Grozin:2023dlk}. The new results of this work are the coefficient $\gamma_2^{(1)}$ and the identification of a  symmetry  relating coefficients at a fixed loop-order with differing powers of $Z$. Entries marked with ``\qMark'' are currently unknown, while entries expressed in terms of other coefficients in the table are fixed by the symmetries discussed in Section~\ref{sec:symm}. \label{tab:summary} }
\end{table}

\section{Discussion \label{sec:disc}}

We have computed the order $Z^2\alpha^3$ coefficient of the anomalous dimension for heavy-heavy-light operators contributing to nuclear beta decay. Our result makes use of simplified Feynman rules that are valid in the static limit of zero nuclear recoil. We have identified new symmetries that relate coefficients in the expansion of the anomalous dimension.
With our new result for $\gamma_{2}^{(1)}$, the first unknown coefficient occurs at $O(Z^2\alpha^4)$ i.e., at four-loop order
and at second sub-leading order in $Z$. 

The combination of the symmetry introduced in Section~\ref{sec:symm} and the leading-$Z$ asymptotics discussed in Ref.~\cite{Hill:2023bfh} implies powerful constraints on logarithmically enhanced contributions to beta decay. As an illustration, we may adopt the ``intermediate-$Z$'' power counting introduced in Ref.~\cite{Hill:2023acw} where $Z\sim L\sim \alpha^{-1/2}$ with $L=\log(\mu_H/\mu_L)$. Using the results summarized in \cref{tab:summary}the running of Wilson coefficients can be computed through $O(\alpha^2)$ with unknown coefficients entering first at $O(Z^2\alpha^4 L) = (\alpha^{5/2})$ and $O(Z^4\alpha^5 L) = O(\alpha^{5/2})$. 

These new perturbative inputs have important numerical impacts for the extraction of $|V_{ud}|$ from superallowed beta decays~\cite{Hardy:2020qwl}. They allow the systematic resummation of logarithms arising from RG evolution from short-distance scales set by nuclear structure, down to long-distance scales set by the reaction's $Q$-value and the electron mass. Given a short-distance matching calculation of $\mathcal{C}(\mu_H)$, and the RG evolution determined by 
Eq.~(\ref{eq:RG}) 
(and discussed in more detail in Ref.~\cite{Hill:2023acw}), 
the remaining operator matrix element can be evaluated perturbatively 
using Eq.~(\ref{eff-L}).  
Evaluation of this matrix element will be presented in future work. 

\vskip 0.1in
\noindent{\bf Acknowledgments}
\vskip 0.1in
Early portions of this work were performed at the Mainz Institute for Theoretical
Physics (MITP) of the DFG Cluster of Excellence PRISMA+ (Project ID
39083149), and RP thanks them for their hospitality. RP thanks the Institute for Nuclear Theory at the University of Washington for its kind hospitality where part of this research was performed, and by extension the INT's U.S. Department of Energy grant No.~DE-FG02-00ER41132. RP is supported by the Neutrino Theory Network under Award Number DEAC02-07CHI11359, the U.S. Department of Energy, Office of Science, Office of High Energy Physics under Award Number DE-SC0011632, and by the Walter Burke Institute for Theoretical Physics. 
This work was supported by the U.S. Department of Energy, Office of Science, Office of High Energy Physics, under Award DE-SC0019095.
R.J.H. acknowledges support from a Fermilab Intensity Frontier Fellowship. 
This manuscript has been authored by Fermi Research Alliance, LLC under Contract No.~DE-AC02-07CH11359 with the U.S. Department of Energy, Office of Science, Office of High Energy Physics.

\pagebreak 
\appendix

\section{Background field Feynman rules \label{sec:backgroundfield}}

Consider the Lagrangian for heavy particles of charge $Q_A=Z+1$ and $Q_B = Z$, interacting by an operator of the form (\ref{eq:Oeff}), 
\begin{align}
    {\cal L} &= \bar{h}_v^{(A)} \left( iv\cdot \partial + e Q_A v\cdot A \right) h_v^{(A)} 
    + 
\bar{h}_v^{(B)} \left( iv\cdot \partial + e Q_B v\cdot A \right) h_v^{(B)} 
+ {\cal C}\, \bar{h}_v^{(B)} \Gamma h_v^{(A)} \, \bar{\nu} \Gamma^\prime e \,,
\end{align}
where ${\cal C} \sim G_F$ is a constant and $v^\mu = (1,0,0,0)$. 
Consider the field redefinition, 
\begin{align}
    h_v^{(A)} = S_v^{(A)} h_v^{(A0)} \,,
\qquad 
    h_v^{(B)} = S_v^{(B)} h_v^{(B0)} \,,
\end{align}
where $S_v^{(A)}$ and $S_v^{(B)}$ are Hermitian 
timelike 
Wilson lines, 
\begin{align}
S_v^{(A)} = \exp\bigg[iZ e \int_{-\infty}^0 ds\, v\cdot A(x+sv) \bigg] \,,
\qquad
S_v^{(B)} = \exp\bigg[- iZ e \int_0^{\infty} ds\, v\cdot A(x+sv) \bigg] \,. 
\end{align}
The choices of integration limits correspond to initial state $A$ particles and final state $B$ particles,
\vspace{5mm}

\begin{align}
\parbox{10mm}{
\begin{fmfgraph*}(30,30)
 \fmftop{t}
 \fmfbottom{b}
 \fmf{photon}{t,b}
  \fmfv{decor.shape=cross}{b}
  \fmflabel{$\mu$}{t}
  \fmflabel{$S_v^{(A)}(0)$}{b}
   \marrow{a}{right}{rt}{$L$}{t,b}
\end{fmfgraph*}
}
\quad &= iZe \, v^\mu \int_{-\infty}^0 ds \, \e^{-isL\cdot v}
\to iZe \, v^\mu {i \over v\cdot L + i 0} \,,
\\[12mm]\nonumber
\parbox{10mm}{
\begin{fmfgraph*}(30,30)
 \fmftop{t}
 \fmfbottom{b}
 \fmf{photon}{t,b}
  \fmfv{decor.shape=cross}{b}
  \fmflabel{$\mu$}{t}
  \fmflabel{$S_v^{(B)\dagger}(0)$}{b}
   \marrow{a}{right}{rt}{$L$}{t,b}
\end{fmfgraph*}
}
\quad &= iZe \, v^\mu \int_0^{\infty} ds \, \e^{-isL\cdot v}
\to
iZe \, v^\mu {i \over -v\cdot L + i 0} \,.
\end{align}
\vspace{8mm}

\noindent
The Lagrangian becomes 
\begin{align}
\label{eq:Lredef}
    {\cal L} &= \bar{h}_v^{(A0)} \left( iv\cdot \partial + e v\cdot A \right) h_v^{(A0)} 
    + 
\bar{h}_v^{(B0)} iv\cdot \partial  h_v^{(B0)} 
+ {\cal C}\, S_v^{(B)\dagger}S_v^{(A)} \,
\bar{h}_v^{(B0)} \Gamma h_v^{(A0)} \, \bar{\nu} \Gamma^\prime e \,,
\end{align}
where the combination of Wilson lines is 
\begin{align}
    S_v^{(B)\dagger}S_v^{(A)}
    &= \exp\bigg[iZ e \int_{-\infty}^\infty ds\,v\cdot A(x+sv) ]
    = \exp\bigg[ 2\pi i Z e \delta(iv\cdot \partial)v\cdot A \bigg] \,.
\end{align}
This Lagrangian leads to the Feynman rules in Fig.~\ref{fig:vertex}. 

We have omitted diagrams in Fig.~\ref{fig:threeloopdiagrams} involving photons 
connecting the background field and the charge $+1$ proton.   
These diagrams correspond to a residual mass for the heavy proton field and do not affect the computation of the anomalous dimension.  Explicitly, 
the contribution of these diagrams to the heavy proton self energy is 
\begin{align}
    -i \,\delta\Sigma_v(k^0) 
    &= \quad
    \parbox{10mm}{
\begin{fmfgraph*}(30,30)
 \fmftopn{t}{3}
 \fmfbottom{b}
 \fmf{photon}{t2,b}
 \fmf{double}{t1,t2,t3}
  \fmfv{decor.shape=cross}{b}
\end{fmfgraph*}
}
\quad 
+
\quad
 \parbox{10mm}{
\begin{fmfgraph*}(30,60)
 \fmftopn{t}{3}
 \fmfbottom{b}
 \fmf{phantom}{t2,b}
 \fmf{double}{t1,t2,t3}
\fmffreeze
\fmf{photon}{t2,a1}
\fmf{photon}{b,a2}
\fmf{fermion,left,tension=0.5}{a1,a2,a1}
  \fmfv{decor.shape=cross}{b}
\end{fmfgraph*}
}
\quad
+ \dots 
\equiv - i \, \delta m \,,
\end{align}
where with photon mass regulator, 
\begin{align}
    \delta m &= { Z e^2 \lambda^{1-2\epsilon} \Gamma\left(-\frac12 + \epsilon\right)
    \over (4\pi)^{\frac32-\epsilon}} + \dots 
    \to -(Z\alpha) \lambda + \dots \,. 
\end{align}
Contributions involving 
$\delta m$ can be removed by the field redefinition $h_v^{(A0)}(x) \to \exp[ i  \, \delta m \, v \cdot x] h_v^{(A0)}(x)$ in Eq.~(\ref{eq:Lredef}).  
The additional diagrams involving $\delta m$ could be evaluated explicitly, yielding finite contributions to amputated amplitudes.  
In this case, an additional regulator (e.g. residual momentum) 
should be used to regulate infrared divergences in diagrams involving multiple heavy proton propagators.  It can be seen explicitly using a momentum regions analysis 
that this additional regulator is unnecessary for the diagrams in Fig.~\ref{fig:threeloopdiagrams}. 

\section{Elementary integrals \label{sec:elementary}}

We collect here results for Euclidean integrals that are used in the amplitude evaluation.  When dimension is not specified, the integration measures refers to dimension $D$:
$(\dd k)=(\dd^Dk)=d^D k/(2\pi)^D$, $(\dd \omega)=d\omega/(2\pi)$. 
First, let us define the elementary integral, 
\begin{align}
    J(a,b) &= \int (\dd \omega) \, (\omega^2)^{a\over 2} (1+\omega^2)^b 
= {1\over 2\pi} \beta\left( {a+1\over 2}, -b-{a+1\over 2} \right)
\,, 
\end{align}
where $\beta(x,y)=\Gamma(x)\Gamma(y)/\Gamma(x+y)$.  
Next, we tabulate some simple one-loop integrals, 
\begin{align}
A(a,m) &= \int (\dd p) {1\over (\bm{p}^2+m^2)^a} = {(m^2)^{\frac{D}{2}-a}\over (4\pi)^{D\over 2}}{\Gamma(a-\frac{D}{2})\over\Gamma(a)} \,, 
\\
L(a,b,m) &= \int (\dd p) {1\over (\bm{p}^2)^a}{1\over (\bm{p}^2+m^2)^b}
 = {(m^2)^{{D\over 2}-a-b}\over (4\pi)^{D\over 2}} 
 {\Gamma(a+b-{D\over 2})\over \Gamma(a)\Gamma(b)} 
\beta\left(a,{D\over 2}-a \right) \,,
\\
M(a,b,m) &= \int (\dd k)(\dd \omega) {1\over (\omega^2+\bm{k}^2)^a}{1\over (\omega^2+\bm{k}^2+m^2)^b} 
= {(m^2)^{{d\over 2}-a-b}\over (4\pi)^{d\over 2}} 
 {\Gamma(a+b-{d\over 2})\over \Gamma(a)\Gamma(b)} 
\beta\left(a,{d\over 2}-a \right)~,
\\
Y(a,b,\bm{q}^2) &= \int (\dd p) {1\over (\bm{p}^2)^{a}}{1\over [(\bm{p}-\bm{q})^2]^{b}} 
= (\bm{q}^2)^{{D\over 2} -a-b} {\Gamma(a+b)\over \Gamma(a)\Gamma(b)} A(a+b,1) \beta\left( {D\over 2}-b, {D\over 2}-a \right) \,.
\end{align}
Another  useful one-loop integral is
\begin{align}
    Z^{(c)}(a,b,p^2) &= \int(\dd \omega)(\dd k) \,
    (\omega^2)^{c\over 2}
    \,{1\over (\omega^2 + \bm{k}^2)^{a}}{1\over (\omega^2 + (\bm{k}+\bm{p})^2)^{b}} 
    \\
    &=(p^2)^{{c+1+D\over 2}-a-b}{\Gamma(a+b)\over \Gamma(a)\Gamma(b)}
    A(a+b,1)J(c,{D\over 2}-a-b) 
    \beta\left( {c+1+D\over 2}-b, {c+1+D\over 2}-a\right)~.\nonumber
\end{align}
Finally, some simple two-loop results often arise as sub-integrals. These are given by
\begin{align}
B(a_1,a_2,a_3,m) &= \int (\dd p)(\dd q) {1\over (\bm{q}^2)^{a_1}}{1\over [(\bm{p}-\bm{q})^2+m^2]^{a_2}}{1\over (\bm{p}^2)^{a_3}}
\\
&= { (m^2)^{{D\over 2}-a_1-a_2-a_3}\over (4\pi)^D}
{\Gamma(a_1+a_2+a_3-D)\over \Gamma(a_1)\Gamma(a_2)\Gamma(a_3)} 
{\beta\left({D\over 2}-a_3,{D\over 2} - a_1\right)\beta\left(D-a_1-a_3,a_1+a_3-{D\over 2} \right) \over \Gamma(a_1+a_3-{D\over 2}) } \,,
\nl
K^{(b)}(a_1,a_2,a_3,m) 
&= \int (\dd \omega)(\dd k)(\dd p)\, (\omega^2)^{b\over 2}
{1\over (\omega^2+\bm{k}^2+m^2)^{a_1}}
{1\over (\bm{p}^2)^{a_2}}
{1\over (\omega^2+(\bm{k}+\bm{p})^2)^{a_3}}
\\
&= { (m^2)^{D-a_1-a_2-a_3 + {1+b\over 2}}\over (4\pi)^D}
{\Gamma(a_1+a_3-{D\over 2})\over \Gamma(a_1)\Gamma(a_3)}
J(b,D-a_1-a_2-a_3) 
L\left(a_2,a_1+a_3-{D\over 2},1\right) \,. \nonumber
\end{align}

\section{Fourier transforms}

The follow Fourier transforms are evaluated in $D=3$: 
\begin{align}
f_n(M,r) &= \int (\dd p) \e^{i\bm{p}\cdot\bm{r}} {1\over (\bm{p}^2+m^2)^n} \,,
\end{align}
with the results (using $r=|\bm{r}|$)
\begin{align}
    f_1(M,r) &= {1\over 4\pi}{\e^{-mr}\over r} \,,
    \\
    f_2(m,r) &= {-1\over 2m} {\partial \over \partial m}f_1(m,r) = {1\over 4\pi}{\e^{-mr}\over 2m} \,, 
    \\
    f_3(m,r) &= {-1\over 4m} {\partial \over \partial m}f_2(m,r) 
    = {1\over 4\pi}{(1+mr)\e^{-mr}\over 8m^3}~.
\end{align}
The following convolution integrals are evaluated in $D=3$ with $p=|\bm{p}|$:
\begin{align}
    g(m_1,m_2,p) &= \int (\dd q) {1\over m_1^2 + \bm{q}^2}{1\over m_2^2 + (\bm{q}+\bm{p})^2 } = {1\over 4\pi p} \arctan\left({p\over m_1+m_2}\right) \,,
    \\
    h(m_1,m_2,p) &= \int (\dd q) {1\over (m_1^2 + \bm{q}^2)^2}{1\over m_2^2 + (\bm{q}+\bm{p})^2 } = {-1\over 2m_1} {\partial \over \partial m_1}g(m_1,m_2,p) 
    =
    {1\over 4\pi} {1\over 2m_1} {1\over {p}^2 + (m_1+m_2)^2} \,.
\end{align}
The following is also useful: 
\begin{align}
    \bm{j}^{(n_1,n_2)}(m_1,m_2,\bm{k}) &= \int(\dd p) {\bm{p}\over (\bm{p}^2+m_1^2)^{n_1}}
    {1\over [m_2^2 + (\bm{p}+\bm{k})^2]^{n_2} }
    = \hat{\bm{k}} \,j^{(n_1,n_2)}(m_1,m_2,k) \,.
\end{align}
We have
\begin{align}
    j^{(1,1)}(m_1,m_2,k) &= {-1\over (4\pi)^2}\bigg[m_1 {\cal A}\left(k,2,{m_1+m_2\over k}\right) + {\cal A}\left(k,3,{m_1+m_2\over k}\right)  \bigg] \,,
\end{align}
with
\begin{align}
    {\cal A}(k,n,\rho) &= k^{n-3} (4\pi)\int_0^\infty dr \, r^{-n+1}\left( \cos{r} - {\sin{r} \over r}\right)\e^{-\rho r} \,,
\end{align}
and then also
\begin{align}
    j^{(2,1)}(m_1,m_2,k) &= {-1\over 2m_1} {\partial \over \partial m_1}j^{(1,1)}(m_1,m_2,k) \,,
    \qquad
    j^{(1,2)}(m_1,m_2,k) = {-1\over 2m_2} {\partial \over \partial m_2}j^{(1,1)}(m_1,m_2,k) \,. 
\end{align}

\section{Amplitudes and numerator algebra}
In this section we describe how to arrive at the integral basis discussed in Eq.~(\ref{eq:Iintegrals}) from the Feynman rules and by use of the projector $\tfrac14 {\rm Tr}[\ldots]$ to simplify numerator algebra. We work in Feynman gauge, however general covariant gauge expressions are easily obtained by
similar manipulations. 
For the diagrams in Fig.~\ref{fig:threeloopdiagrams}, 
we label the momentum flowing through the outer-most Coulomb photon by $\vb{q}$, the momentum flowing through the inner-most Coulomb photon by $\vb{p}-\vb{q}$ and the momentum flowing through the ``dynamical'' photon (i.e., that with non-zero energy transfer) by $k_\mu = (k_0, \vb{k})$.  Let us introduce the denominators, 
\begin{align}
    D_1 &= \omega^2+(\vb{p}+\vb{k})^2 ~, \\
    D_2 &= \vb{p}^2 ~,\\
    D_3 &= (\vb{p}-\vb{q})^2+\lambda^2 ~,\\
    D_4 &= \omega^2+(\vb{q}+\vb{k})^2 ~, \\
    D_5 &= \vb{q}^2 ~,\\
    D_6 &= \omega^2+\vb{k}^2 ~,\\
    D_7 &= \vb{q}^2+\lambda^2 ~,\\
    D_8 &= \omega^2+ \vb{k}^2+\lambda^2~.
\end{align}
For the numerator algebra it is convenient to introduce 
four-vectors $q_\mu=(0,\vb{q})$ such that $\slash{q}=q^\mu\gamma_\mu = - \vb*{\gamma}\cdot \vb{q}$.  Similarly, $p_\mu=(0,\vb{p})$. 

Let us now use the Feynman rules in Figs.~\ref{fig:vertex} and \ref{fig:prop} to determine 
the relevant integrals that must be computed. 
Consider e.g. $\mathcal{M}^w$ from Fig.~\ref{fig:threeloopdiagrams}.
Direct evaluation using the Feynman rules and Wick rotation $k_0 \rightarrow  i  \omega$ gives
\begin{equation}    
    \label{app_eq:Mw_1}
    \mathcal{M}^w=  - Z^2 e^6 \int (\dd p) (\dd q)   (\dd k)(\dd \omega) \frac{\slash{p}}{D_2}\gamma_\mu  
    \frac{\slash{p}+\tilde{\slash{k}}}{D_1} \gamma_0 \frac{\slash{q}+  \tilde{\slash{k}}}{D_4}\gamma_0 \frac{\tilde{\slash{k}}}{D_6}\gamma^\mu \frac{1}{D_3} \frac{1}{D_7} \frac{1}{D_8}  \,,
\end{equation}
where we define
 $\tilde{\slash{k}}= i  \omega \gamma_0 -\vb*{\gamma}\cdot \vb{k}$.  Acting with $\frac14{\rm Tr}[\ldots]$ we find, 
\begin{equation}    
    \label{app_eq:Mw_2}
    \mathcal{M}^w=  Z^2 e^6 \int (\dd q') (\dd q)   (\dd k)(\dd \omega) \frac{(1-\epsilon) [D_1 (D_5-D_4)+D_6 (D_7-D_3+D_4-D_5)+4 \omega ^2 (D_1-D_6)] }{D_1 D_2 D_3 D_4 D_6 D_7 D_8} ~.
\end{equation}
Using partial fraction identities this can be re-expressed in terms of the integral basis presented in the main text.

Similarly for diagrams $(a)$, $(b)$, $(c)$, $(p1)$ and $(p2)$, 
direct evaluation using the Feynman rules, Wick rotating, acting with $\frac14{\rm Tr}[\ldots]$ and performing numerator algebra gives
\begin{align}   
    \begin{split}
    \label{app_eq:Ma}
    \mathcal{M}^a&=  Z^2 e^6 \int (\dd p) (\dd q)   (\dd k)(\dd \omega) \frac{1}{i\omega} 
    \frac{\slash{p}+\tilde{\slash{k}}}{D_4}\gamma_0\frac{\slash{p}}{D_2}\gamma_0 \frac{\slash{q}} {D_5}   \gamma_0 
      \frac{1}{D_3} \frac{1}{D_7} \frac{1}{D_8}~\\
    &=  Z^2 e^6 \int (\dd q') (\dd q)   (\dd k)(\dd \omega) \frac{1}{2}\frac{ \left(D_2-D_3+D_7\right) }{D_2 D_3 D_4 D_5 D_7 D_8} \,,
    \end{split}
    \\
    \begin{split}
    \label{app_eq:Mb}
    \mathcal{M}^b&=   Z^2 e^6 \int (\dd p) (\dd q)   (\dd k)(\dd \omega) \frac{1}{i\omega}  
    \frac{\slash{p}+\tilde{\slash{k}}}{D_1}\gamma_0  \frac{\slash{q}+ \tilde{\slash{k}}}{D_4} \gamma_0 \frac{\slash{q}} {D_5}\gamma_0
    \frac{1}{D_3} \frac{1}{D_7} \frac{1}{D_8}  ~\\
    &=  Z^2 e^6 \int (\dd q') (\dd q)   (\dd k)(\dd \omega) \frac{1}{2} \frac{\left( D_2-D_3+2D_4-2D_6+D_7 \right) }{D_1 D_3 D_4 D_5  D_7 D_8} \,,
    \end{split}
    \\
    \begin{split}
    \label{app_eq:Mc}
    \mathcal{M}^c&=  Z^2 e^6 \int (\dd p) (\dd q)   (\dd k)(\dd \omega) \frac{1}{i\omega} 
    \frac{\slash{p}+\tilde{\slash{k}}}{D_1}\gamma_0\frac{\slash{q}+ \tilde{\slash{k}}}{D_4} \gamma_0  \frac{\tilde{\slash{k}}} {D_6}\gamma_0 
    \frac{1}{D_3} \frac{1}{D_7} \frac{1}{D_8}  ~\\
    &=  Z^2 e^6 \int (\dd q') (\dd q)   (\dd k)(\dd \omega) \frac{1}{2} \frac{\left( 2 D_1-D_2-D_3+2D_4-2D_5+2D_6+D_7-8 \omega^2\right) }{ D_1 D_3 D_4 D_6  D_7 D_8} \,,
    \end{split}
    \\  
    \begin{split}
    \label{app_eq:Mp1}
    \mathcal{M}^{p1}&=  (-1) Z^2 e^6 \int (\dd p) (\dd q)   (\dd k)(\dd \omega)  
    \frac{\slash{p}}{D_2}\gamma_0\frac{\slash{q}}{D_5} \gamma^\mu\frac{\slash{q}+ \tilde{\slash{k}}}{D_4} \gamma_\mu  \frac{\slash{q}} {D_5} \gamma_0
    \frac{1}{D_3} \frac{1}{D_7} \frac{1}{D_8}  ~\\
    &=  Z^2 e^6 \int (\dd q') (\dd q)   (\dd k)(\dd \omega) \frac{(1-\epsilon)\left[ D_1 (D_3-D_2-D_7)+D_2 (D_4-D_5)+D_6 (D_7-D_3)\right] }{D_1 D_2^2 D_3 D_5  D_7 D_8 } \,,
    \end{split}
    \\
    \label{app_eq:Mp2}
    \mathcal{M}^{p2}&=  (-1) Z^2 e^6 \int (\dd p) (\dd q)   (\dd k)(\dd \omega) 
     \frac{\slash{p}}{D_2} \gamma^\mu \frac{\slash{p}+\tilde{\slash{k}}}{D_1}\gamma_\mu \frac{\slash{p}}{D_2}\gamma_0  \frac{\slash{q}} {D_5}\gamma_0
    \frac{1}{D_3} \frac{1}{D_7} \frac{1}{D_8}  ~\\
    &=  Z^2 e^6 \int (\dd q') (\dd q)   (\dd k)(\dd \omega)  \frac{ (1-\epsilon) [D_1 D_5- D_2 (D_4+D_5-D_6)- D_3 (D_6-D_4)-D_4 D_7 -D_5 D_6+ D_6 D_7] }{D_2 D_3 D_4 D_5^2 D_7 D_8 } ~.\nonumber
\end{align}
Vertex correction diagrams $(v1)$ and $(v2)$ involve substantially more complicated numerator algebra. 
Following the same steps as above we have
\begin{align}
    \mathcal{M}^{v1}&= Z^2 e^6 \int (\dd p) (\dd q)   (\dd k)(\dd \omega)  \frac{N_0^{v1} + \epsilon N_1^{v1}}{D_1 D_2 D_3D_4D_5D_7D_8} \,, \\
    \mathcal{M}^{v2}&= Z^2 e^6 \int (\dd p) (\dd q)   (\dd k)(\dd \omega)  \frac{N_0^{v2} + \epsilon N_1^{v2}}{D_2 D_3 D_4 D_5 D_6 D_7D_8} \,.
\end{align}
with 
\begin{align}
    N_0^{v1}&= -D_1 (-D_2+D_3+D_5-D_7)-D_2 (D_3+D_5-D_7+2\omega^2)+D_3^2-D_3 D_4+D_3 D_5+D_3 D_6 \\
            &~~~~~~- 2 D_3 D_7+2 D_3 \omega^2+D_4 D_7+D_5 D_6-D_5 D_7-D_6 D_7+D_7^2-2 D_7 \omega ^2~, \nonumber \\
    N_1^{v1}&= -D_1 D_5-D_2 (D_4-2\omega^2)+D_3 D_6-2 D_3 \omega^2+D_5 D_6-D_6 D_7+2 D_7 \omega^2,\\
    N_0^{v2}&=-D_1 D_5+D_2 D_4-2 D_2\omega^2-D_3 D_4+D_3 D_5+2 D_3 \omega^2+D_4 D_7+D_5 D_6-D_5 D_7-2 D_7 \omega^2~,\\
    N_1^{v2}&=-D_1 D_5+2\omega^2 (D_2-D_3+D_7)+D_2 D_5-D_2 D_6+D_3 D_6+D_5 D_6-D_6 D_7~.
\end{align}
The evaluation of $\mathcal{M}^{b1}$ and $\mathcal{M}^{b2}$ is discussed explicitly in Sec.~\ref{bubble-eval}. Having obtained explicit integral representations of the amplitudes, Appendix~\ref{sec:IBP} maps to the integral basis of Eq.~(\ref{eq:Iintegrals}) and describes the evaluation of the integrals.

\section{Amplitude expansion \label{sec:IBP}}

Let us consider the amplitudes corresponding to the diagrams in Fig.~\ref{fig:threeloopdiagrams}, in terms of the basis of integrals (\ref{eq:Iintegrals}). 
Gauge dependence affects only the photon propagator that is not connected to the background field, and the general amplitude has the structure 
\begin{align}
 {\cal M} =  {\cal M}\big|_{\xi=1} + (1-\xi) \Delta {\cal M} \,.
\end{align}
Subamplitudes ${\cal M}^{a,b,c}\big|_{\xi=1}$ have 
been given above and corresponding gauge-dependent pieces are 
\begin{align}
    \Delta {\cal M}^a &= Z^2 e^6  \bigg\{
    \frac12
\bigg[- \bm{1}^- \bm{2}^- 
+\bm{1}^- \bm{3}^- 
-\bm{1}^- \bm{5}^-
+ (\bm{2}^-)^2 
- \bm{2}^- \bm{3}^-
+ \bm{2}^- \bm{4}^-
- \bm{2}^- \bm{6}^-
\bigg]
I_{111011}
+ \delta \Delta I^a 
    \bigg\} \,,
    \nl
    \Delta {\cal M}^b &= 
    Z^2 e^6 \bigg\{ 
    -2 I^{(2)}_{101101} + \frac12\bigg[ 
    \bm{1}^- \bm{5}^- 
    -\bm{2}^- \bm{4}^-
    +\bm{3}^- \bm{4}^- 
    -\bm{3}^- \bm{5}^-
    -(\bm{4}^-)^2
    +\bm{4}^- \bm{5}^- 
    +\bm{4}^- \bm{6}^-
    \bigg]I_{101111}
    + \delta\Delta I^b
    \bigg\}
    \,,
    \nl
    \Delta {\cal M}^c &=
    Z^2 e^6 \bigg\{ 
    2 I^{(2)}_{101101}
    + \frac12 \bigg[
    -\bm{1}^- \bm{6}^-
    + \bm{3}^- \bm{6}^- 
    - \bm{4}^- \bm{6}^- 
    \bigg] I_{101102}
    + \delta \Delta I^c
    \bigg\} \,.
\end{align}
The quantities $\delta \Delta I^{a,b,c}$ 
refer to the replacement (\ref{eq:deltareplace}) and
can be evaluated by isolating subdivergences, yielding
\begin{align}
    \delta \Delta I^a &= 
    \left[(4\pi)^\epsilon \Gamma(1+\epsilon) \over (4\pi)^2\right]^3 \left( \frac{8\pi^2 }{\epsilon} \right) \,, \quad
    \delta \Delta I^b = \delta \Delta I^c = 0 \,. 
\end{align}
Amplitudes ${\cal M}^{b1}\big|_{\xi=1}$ and ${\cal M}^{b2}\big|_{\xi=1}$ have been evaluated in Eqs.~(\ref{eq:b1}) and (\ref{eq:b2}), and 
$\Delta {\cal M}^{b1} = \Delta {\cal M}^{b2} = 0$.
For the remaining subamplitudes, we have, at $\xi=1$, 
\begin{align}
     {\cal M}^{p1}\big|_{\xi=1}&=Z^2 e^6 \bigg\{
(1-\epsilon) \bigg[ \bm{1}^- \bm{2}^- 
-\bm{1}^- \bm{3}^-
+\bm{1}^- \bm{5}^- 
-\bm{2}^- \bm{4}^-
+\bm{2}^- \bm{5}^- 
+\bm{3}^- \bm{6}^-
-\bm{5}^- \bm{6}^-
\bigg]I_{121010}
+ \delta I^{p1}
     \bigg\}
\,,    \nl
     {\cal M}^{p2}\big|_{\xi=1}&= 
     Z^2 e^6 \bigg\{
     (1-\epsilon)\bigg[
      -\bm{1}^-\bm{5}^- 
     +\bm{2}^- \bm{4}^-
     +\bm{2}^- \bm{5}^-
     -\bm{2}^- \bm{6}^-
     -\bm{3}^- \bm{4}^-
     +\bm{3}^- \bm{6}^-
     +\bm{4}^- \bm{5}^-      
     \bigg]I_{011120} + \delta I^{p2}
     \bigg\}
\,,    \nl
     {\cal M}^{v1}\big|_{\xi=1}&= 
     Z^2 e^6 \bigg\{
    2(1-\epsilon)\big[ I^{(2)}_{101110} -I^{(2)}_{110110} +I^{(2)}_{111100}  \big]
    + \big[ \epsilon\big( \bm{1}^-\bm{5}^- + \bm{2}^- \bm{4}^- \big)
    -(1+\epsilon)\bm{3}^- \bm{6}^-
    \nl
    &\quad 
    - \bm{1}^- \bm{2}^- 
    + \bm{1}^- \bm{3}^- 
    + \bm{2}^- \bm{3}^- 
    -(\bm{3}^-)^2 
    + \bm{3}^- \bm{4}^- 
    + \bm{3}^- \bm{5}^-
    -\bm{4}^- \bm{5}^- 
    \big]I_{111110}
    + \delta I^{v1}
     \bigg\}
\,,    \nl
     {\cal M}^{v2}\big|_{\xi=1}&= Z^2 e^6 \bigg\{
2(1-\epsilon)\big[ I^{(2)}_{001111} - I^{(2)}_{010111}+ I^{(2)}_{011101} \big]
+ \big[ \epsilon\big( -\bm{2}^-\bm{5}^- + \bm{2}^- \bm{6}^- -\bm{3}^- \bm{6}^- \big) 
+ (1+\epsilon) \bm{1}^- \bm{5}^- 
\nl &\quad 
-\bm{2}^- \bm{4}^- 
+ \bm{3}^- \bm{4}^- 
-\bm{3}^- \bm{5}^- 
-\bm{4}^- \bm{5}^-
+ (\bm{5}^-)^2 
-\bm{5}^- \bm{6}^- 
\big]I_{011111} + \delta I^{v2}
\bigg\}
\,,    \nl
     {\cal M}^{w}\big|_{\xi=1}&=  Z^2 e^6 \bigg\{
4(1-\epsilon) \big[  -I^{(2)}_{011101} + I^{(2)}_{111100}  \big]
+ (1-\epsilon) \big[ \bm{1}^- \bm{4}^- -\bm{1}^- \bm{5}^- +\bm{3}^- \bm{6}^- -\bm{4}^- \bm{6}^-  
\big]I_{111101} + \delta I^w
     \bigg\}
     \,,
\end{align}
with 
\begin{align}
    \delta I^{p1} + \delta I^{p2}+ \delta I^{v1} + \delta I^{v2} = 0 \,, \quad
    \delta I^w = 0 \,.
\end{align}
For the remaining gauge-dependent terms, 
\begin{align}
     \Delta{\cal M}^{p1}&= 
      Z^2 e^6 \bigg\{
\bigg[ \bm{1}^- \bm{2}^- \bm{3}^- 
-\bm{1}^- (\bm{2}^-)^2  
+ \frac12\bigg(
-\bm{1}^- \bm{2}^- \bm{5}^- 
- \bm{1}^- \bm{2}^- \bm{4}^- 
+ \bm{1}^- \bm{3}^- \bm{6}^- 
-\bm{1}^- \bm{5}^- \bm{6}^- 
\nl
&\quad 
+ (\bm{2}^-)^3
- (\bm{2}^-)^2 \bm{6}^-
- (\bm{1}^-)^2 \bm{3}^-
+ (\bm{1}^-)^2 \bm{5}^- 
+ (\bm{1}^-)^2 \bm{2}^- 
- (\bm{2}^-)^2 \bm{3}^-
+ (\bm{2}^-)^2 \bm{4}^-
\bigg) \bigg]I_{121011}
+ \delta \Delta I^{p1}
     \bigg\}
\,,    \nl
     \Delta{\cal M}^{p2}&= Z^2 e^6 \bigg\{
     \bigg[
\bm{3}^- \bm{4}^- \bm{5}^- 
-\bm{4}^- (\bm{5}^-)^2 
+ \frac12\bigg( 
-\bm{2}^-\bm{4}^-\bm{6}^-
+\bm{3}^-\bm{4}^-\bm{6}^- 
-\bm{1}^-\bm{4}^-\bm{5}^- 
-\bm{2}^-\bm{4}^-\bm{5}^- 
\nl &\quad 
+(\bm{5}^-)^3 
-\bm{3}^- (\bm{4}^-)^2 
+(\bm{4}^-)^2 \bm{5}^-
+\bm{2}^- (\bm{4}^-)^2 
-\bm{3}^- (\bm{5}^-)^2
-\bm{6}^- (\bm{5}^-)^2 
+\bm{1}^- (\bm{5}^-)^2 
\bigg) 
\bigg] I_{011121}
+ \delta \Delta I^{p2} 
     \bigg\}
\,,    \nl
     \Delta{\cal M}^{v1}&= Z^2 e^6 \bigg\{
-2 I^{(2)}_{101101}
+ \frac12 \bigg[ 
\bm{1}^- \bm{2}^- \bm{5}^- 
+ \bm{1}^- \bm{2}^- \bm{4}^- 
+ \bm{1}^- \bm{5}^- \bm{6}^- 
+ \bm{2}^- \bm{4}^- \bm{6}^- 
+ \bm{1}^- \bm{4}^- \bm{5}^- 
+ \bm{2}^- \bm{4}^- \bm{5}^- 
- \bm{1}^- \bm{3}^- \bm{4}^- 
\nl &\quad
+ \bm{2}^- \bm{3}^- \bm{4}^- 
+ \bm{1}^- \bm{3}^- \bm{5}^- 
- \bm{2}^- \bm{3}^- \bm{5}^- 
- \bm{2}^- (\bm{4}^-)^2 
- \bm{1}^- (\bm{5}^-)^2
- (\bm{1}^-)^2 \bm{5}^- 
- (\bm{2}^-)^2 \bm{4}^- 
\bigg]I_{111111} 
+ \delta \Delta I^{v1}
     \bigg\}
\,,    \nl
     \Delta{\cal M}^{v2}&= Z^2 e^6 \bigg\{ 
\frac12 \bigg[ 
    -\bm{1}^- \bm{5}^- \bm{6}^- 
    +\bm{2}^- \bm{4}^- \bm{6}^- 
    -\bm{3}^- \bm{4}^- \bm{6}^- 
    + \bm{3}^- \bm{5}^- \bm{6}^- 
    + \bm{4}^- \bm{5}^- \bm{6}^- 
    - \bm{6}^- (\bm{5}^-)^2 
    + \bm{5}^- (\bm{6}^-)^2 
\bigg]I_{011112}
\nl &\quad
+ \delta \Delta I^{v2} 
     \bigg\} 
\,,    \nl
     \Delta{\cal M}^{w}&=  Z^2 e^6 
     \bigg\{ 
    2 I^{(2)}_{101101} + \frac12 \bigg[ 
    (\bm{1}^-)^2 \bm{6}^- 
    -\bm{1}^- \bm{2}^- \bm{6}^- 
    -\bm{1}^- \bm{3}^- \bm{6}^- 
    + \bm{1}^- \bm{5}^- \bm{6}^- 
    -\bm{1}^- (\bm{6}^-)^2 
    +\bm{2}^-\bm{3}^- \bm{6}^- 
    \nl &\quad
    - \bm{2}^- \bm{4}^- \bm{6}^- 
    \bigg]I_{111102} 
    + \delta \Delta I^w
     \bigg\}
     \,,
\end{align}
with 
\begin{align}
    \delta \Delta I^{p1} + \delta \Delta I^{p2}+ \delta \Delta I^{v1} + \delta \Delta I^{v2} = 0 \,, \quad
    \delta \Delta I^w = 0 \,.
\end{align}

The necessary integrals $I^{(b)}_{a_1 a_2 a_3 a_4 a_5 a_6}$ can be evaluated as follows. 
For cases involving a single difference of the momenta $\bm{k}$, $\bm{p}$ and $\bm{q}$ in Eq.~(\ref{eq:Iintegrals}), the integrals factorize and can be evaluated using the elementary integrals in Appendix~\ref{sec:elementary}.  
For example, the first term in Eq.~(\ref{eq:Maexpand}) for ${\cal M}^{(a)}$ 
is 
\begin{align}\label{eq:Ma1}
    I_{110010} &= \bigg[\int(\dd \omega)(\dd k)(\dd p) {1\over \omega^2 + (\bm{k}+\bm{p})^2}{1\over \bm{p}^2}{1\over \omega^2 + \bm{k}^2 + \lambda^2}\bigg]
    \bigg[ \int (\dd q) {1\over \bm{q}^2 (\bm{q}^2+\lambda^2)}\bigg]
    = L(1,1,\lambda) K(1,1,1,\lambda) 
    \sim 0 \,. 
\end{align}
Integrals involving two momentum differences can be evaluated by first observing that for these single-scale integrals,
\begin{align}
 \lambda {\dd\over \dd \lambda} 
 I^{(b)}_{a_1 a_2 a_3 a_4 a_5 a_6}
 = [3(D-1)+b-2(a_1+a_2+a_3+a_4+a_5+a_6)]\,
I^{(b)}_{a_1 a_2 a_3 a_4 a_5 a_6} \,.
\end{align}
After isolating and subtracting any divergent subintegrals, the remaining integral has $1/\epsilon$ divergence given by convolution integrals in $D=3$. 
For example, the second term in 
Eq.~(\ref{eq:Maexpand}) for ${\cal M}^{(a)}$ is 
\begin{align}\label{eq:Ma2}
    I_{101010}&= {1\over 3\epsilon} 
    \left( -\frac12 \lambda {d\over d\lambda}\right) I_{101010}
    \nl
    &={1\over 3\epsilon}
    \int(\dd \omega)(\dd k)(\dd p)(\dd q)\bigg\{ {1\over \omega^2 + (\bm{k}+\bm{p})^2}{1\over (\bm{p}-\bm{q})^2}{1\over \bm{q}^2(\bm{q}^2+\lambda^2)}{\lambda^2 \over (\omega^2+\bm{k}^2+\lambda^2)^2}
    \nl
    &\qquad
    +
    {1\over \omega^2 + (\bm{k}+\bm{p})^2}\bigg[{1\over (\bm{p}-\bm{q})^2} - {1\over\bm{p}^2}\bigg]
    {\lambda^2\over \bm{q}^2(\bm{q}^2+\lambda^2)^2}{1\over \omega^2+\bm{k}^2+\lambda^2} \bigg\}
    \nl
    &\quad
    + {1\over 3\epsilon}
    \int(\dd \omega)(\dd k)(\dd p)(\dd q) 
    {1\over \omega^2 + (\bm{k}+\bm{p})^2}{1\over \bm{p}^2}
    {\lambda^2\over \bm{q}^2(\bm{q}^2+\lambda^2)^2}{1\over \omega^2+\bm{k}^2+\lambda^2} 
    \nl
    &\sim 
    {1\over 3\epsilon}
    \int {d\omega \over 2\pi} \int d^3r\, 
    \Bigg\{ \lambda^2 f_2(\sqrt{\omega^2+\lambda^2},r)f_1(\omega,r) f_1(0,r) {1\over \lambda^2}\big[ f_1(0,r)- f_1(\lambda,r)\big]
    \nl
    &\quad 
    + f_1(\sqrt{\omega^2+\lambda^2},r)f_1(\omega,r)f_1(0,r) \bigg\{
{1\over \lambda^2}\big[ f_1(0,r) - f_1(\lambda,r) \big] - f_2(\lambda,r) 
\nl
&\quad 
    - \lim_{r\to 0}\bigg[ {1\over \lambda^2}\big[ f_1(0,r) - f_1(\lambda,r) \big] - f_2(\lambda,r) \bigg]
    \bigg\} \Bigg\}
    + {1\over 3\epsilon} \lambda^2 L(1,2,\lambda) K(1,1,1,\lambda) 
     \nl
    &\sim \left[(4\pi)^\epsilon \Gamma(1+\epsilon) \over (4\pi)^2\right]^3
    \left( -{8\pi^2\over 3\epsilon} \right)  \,.
\end{align}
Similarly, the third term in Eq.~(\ref{eq:Maexpand}) for ${\cal M}^{(a)}$ is 
\begin{align}\label{eq:Ma3}
    I_{111000} &= 
    {1\over 3\epsilon} \lambda^2 B(1,2,1,\lambda) M(1,1,\lambda) 
    \nl
    &\quad
    + {1\over 3\epsilon} \int (\dd \omega)(\dd p){1\over \bm{p}^2}
    \bigg[ \int (\dd k) {\lambda^2 \over (\omega^2 + \bm{k}^2+\lambda^2)^2}{1\over \omega^2 + (\bm{k}+\bm{p})^2}\bigg]
    \bigg[ \int (\dd q) {1\over \bm{q}^2 + \lambda^2}{1\over (\bm{p}-\bm{q})^2} \bigg]
    \nl
    &\quad 
     + {1\over 3\epsilon} \int (\dd \omega)(\dd p){1\over \bm{p}^2}
    \bigg[ \int (\dd k) {1 \over \omega^2 + \bm{k}^2+\lambda^2}{1\over \omega^2 + (\bm{k}+\bm{p})^2}
    - \lim_{p\to 0} 
     \int (\dd k) {1 \over \omega^2 + \bm{k}^2+\lambda^2}{1\over \omega^2 + (\bm{k}+\bm{p})^2}
    \bigg]
    \nl
    &\qquad\qquad  
    \bigg[ \int (\dd q) {\lambda^2\over (\bm{q}^2 + \lambda^2)^2}{1\over (\bm{p}-\bm{q})^2} \bigg]
    \nl
    &=  {1\over 3\epsilon} \lambda^2 B(1,2,1,\lambda) M(1,1,\lambda) 
    \nl
    &\quad
    + \int (\dd \omega)(\dd p){1\over \bm{p}^2} 
    \bigg\{ \lambda^2 h(\sqrt{\omega^2+\lambda^2},\omega,p)g(\lambda,0,p)
    + \big[ g(\sqrt{\omega^2+\lambda^2},\omega,p)-g(\sqrt{\omega^2+\lambda^2},\omega,0) \big]
    \lambda^2 h(\lambda,0,p) \bigg\} 
     \nl
    &\sim \left[(4\pi)^\epsilon \Gamma(1+\epsilon) \over (4\pi)^2\right]^3
    \left( {8\pi^2\over 3\epsilon^2} + {32\pi^2 \over 3\epsilon} \right) \,.
\end{align}
Using Eq.~(\ref{eq:Maexpand}), together with the explicit expressions (\ref{eq:deltaIa}),  (\ref{eq:Ma1}), (\ref{eq:Ma2}) and (\ref{eq:Ma3}), 
yields ${\cal M}^{(a)}\big|_{\xi=1}$ in Eq.~(\ref{eq:Msummary}). 

The remaining amplitudes are evaluated similarly.  
Some basis integrals involve all three momentum differences (i.e., $a_1\ne 0$, $a_3 \ne 0$, $a_4\ne 0$), but can be written as integrals involving only two momentum differences by change of variable, and then evalutated by the above procedure.  For example, from ${\cal M}^{(b)}$, with $\bm{k} = \bm{k}^\prime + \bm{q}^\prime$, $\bm{p} = \bm{p}^\prime - \bm{q}^\prime$, $\bm{q} = -\bm{q}^\prime$, 
\begin{align}
    I_{101100} &= 
    \int (\dd \omega)(\dd k^\prime)(\dd p^\prime)(\dd q^\prime) 
    {1\over \omega^2 + \bm{k}^{\prime 2}}{1\over \bm{p}^{\prime 2}}{1\over \omega^2 + (\bm{k}^\prime + \bm{p}^\prime)^2}
    {1\over \bm{q}^{\prime 2} + \lambda^2}
    {1\over \omega^2 + (\bm{k}^\prime + \bm{q}^\prime)^2 + \lambda^2}
    \nl 
    &\sim {1\over 3\epsilon} \int (\dd \omega)(\dd k^\prime) {1\over \omega^2+\bm{k}^{\prime 2}}g(0,\omega,k^\prime)\bigg[ \lambda^2 h(\lambda,\sqrt{\omega^2+\lambda^2},k^\prime) + \lambda^2 h(\sqrt{\omega^2+\lambda^2},\lambda,k^\prime) \bigg]
\,.
\end{align}
Some integrals involve all three momentum differences (i.e., $a_1\ne 0$, $a_3 \ne 0$, $a_4\ne 0$), but one of these appears in a numerator.  
For example, (after change of variable) 
\begin{align}\label{eq:I1m11110}
    I_{1-11110} &= 
    \int (\dd \omega) (\dd k^\prime) {1\over \omega^2 + \bm{k}^{\prime 2}}
    \bigg[  \int (\dd p^\prime) {1\over \bm{p}^{\prime 2}} {1\over \omega^2 + (\bm{k}^{\prime} + \bm{p}^\prime)^2}  \bigg]
    \bigg[  \int (\dd q^\prime) 
    {1\over \bm{q}^{\prime 2} (\bm{q}^{\prime 2} + \lambda^2)}
    {1\over \omega^2 + (\bm{k}^\prime + \bm{q}^\prime)^2 + \lambda^2}
     \bigg]
     \nl &\quad 
     (\bm{p}^\prime - \bm{q}^\prime)^2 
     \nl 
    &\sim I_{100110} + I_{101100} 
    \nl
    &\qquad 
    -2 \int (\dd \omega)(\dd k^\prime) {1\over \omega^2 + k^{\prime 2}} j^{(1,1)}(0,\omega,k^\prime) {1\over \lambda^2}\bigg[ j^{(1,1)}(0,\sqrt{\omega^2+\lambda^2},k^\prime) - j^{(1,1)}(\lambda,\sqrt{\omega^2+\lambda^2},k^\prime) \bigg]
  \,. 
\end{align}
An exception is $I^{(2)}_{111100}$.  This integral can be evaluated by noting either of the integration by parts (IBP) identities, 
\begin{align}
 D -(a_1+a_2 + 2 a_3) &= a_1 \bm{1}^+ (\bm{3}^- - \bm{4}^-) + a_2 \bm{2}^+ (\bm{3}^- - \bm{5}^-) \,,
 \nl
 D - (a_1 + 2a_2 + a_3) &= a_1 \bm{1}^+ (\bm{2}^- - \bm{6}^-) + a_3 \bm{3}^+ (\bm{2}^- - \bm{5}^-) \,, 
\end{align}
which result by using that in dimensional regularization
\begin{align}
0 &= \int (\dd \omega)(\dd k)(\dd p)(\dd q) {\partial \over \partial p^i} (p^i - q^i) 
\, {\cal I}(\omega,\bm{k},\bm{p},\bm{q}) \,,
\nl
0 &= \int (\dd \omega)(\dd k)(\dd p)(\dd q) {\partial \over \partial p^i} p^i 
\, {\cal I}(\omega,\bm{k},\bm{p},\bm{q})  \,, 
\end{align}
for functions ${\cal I}$ given by the integrands in Eq.~(\ref{eq:Iintegrals}). 
Using these relations, we have
\begin{align}\label{eq:IBP}
    I^{(2)}_{111100} &= {1 \over D-4}\bigg[ I^{(2)}_{210100} - I^{(2)}_{211000}+I^{(2)}_{120100}-I^{(2)}_{1211-10}  \bigg] \,,
    \nl
    I^{(2)}_{111100}
    &= {1\over D-4} \bigg[ I^{(2)}_{201100}-I^{(2)}_{21110-1}+I^{(2)}_{102100}- I^{(2)}_{1121-10} \bigg] \,.
\end{align}
The  basis integrals on the right hand side of Eq.~(\ref{eq:IBP}) 
contain at most two differences of momenta, or a negative subscript $a_5=-1$ or $a_6=-1$.  
The former integrals are evaluated as above.  
The latter integrals simplify after partial fractioning. 
For example, using the second IBP identity for $I^{(2)}_{111100}$ in Eq.~(\ref{eq:IBP}), we require
\begin{align}
    I^{(2)}_{21110-1} &\sim Z^{(2)}(1,2,1) B(3-d/2,1,1,\lambda) 
    \,, 
\end{align}
and 
\begin{align}
I^{(2)}_{1121-10} &\sim 
-\lambda^2 B(1,1,2,\lambda) {1\over d} M(1,1,\lambda) 
\nl
& + \int (\dd \omega)(\dd k^\prime)(\dd p^\prime)(\dd q^\prime) {\omega^2 \over \omega^2 + \bm{k}^{\prime 2}}{1\over \bm{p}^{\prime 2}}{1\over [ (\bm{p}^\prime -\bm{q}^\prime)^2]^2}\bigg[{1\over \omega^2 + (\bm{k}^\prime + \bm{q}^\prime)^2} - {1\over \omega^2 + \bm{k}^{\prime 2}} \bigg]{1\over \omega^2 + (\bm{k}^\prime + \bm{p}^\prime)^2 +\lambda^2} \nl
&\sim -\lambda^2 B(1,1,2,\lambda) {1\over d} M(1,1,\lambda) 
\nl
&\qquad 
+ {1\over 3\epsilon}\int (\dd \omega)(\dd k){\omega^2 \over \omega^2 + k^2}
h(\sqrt{\omega^2+\lambda^2},0,k) \lim_{\delta\to 0} 
\bigg[  h(\delta,\omega,k)- {1\over \omega^2 + k^2} f_2(\delta,0) \bigg] \,.
\end{align}
The integration by parts identities (\ref{eq:IBP}) also provide an alternative evaluation of other integrals; for example, $I_{1-11110}$ in Eq.~(\ref{eq:I1m11110}) can be evaluated using the first identity in Eq.~(\ref{eq:IBP}). 

\end{fmffile} 

\vfill
\pagebreak

\bibliography{Z2a3}

\end{document}